\documentclass[trackchanges, astrosymb, twocolumn]{aastex701}

\usepackage{graphicx}
\usepackage{txfonts}
\usepackage{xcolor}
\usepackage{soul}
\usepackage{comment}
\DeclareRobustCommand{\VAN}[3]{#2}
\let\VANthebibliography\thebibliography
\def\thebibliography{\DeclareRobustCommand{\VAN}[3]{##3}\VANthebibliography}

\newcommand{\msun}{\mathrm{M}_{\odot}}
\newcommand{\mstar}{M_{\star}}

\defcitealias{wen2024}{WH24}
\defcitealias{yang2021}{Y21}
\defcitealias{Dey2019}{DLIS}
\defcitealias{gwyn2025}{UNIONS}

\begin{document}

\title{Cluster properties as a function of dynamical state in the DESI Legacy x UNIONS surveys}

\author[orcid=0000-0001-6336-642X,sname='Ahad']{Syeda Lammim Ahad}
\affiliation{Waterloo Centre for Astrophysics, University of Waterloo, Waterloo, ON N2L 3G1, Canada}
\affiliation{Department of Physics and Astronomy, University of Waterloo, 200 University Avenue West, Waterloo, Ontario N2L 3G1, Canada}
\affiliation{Center for Astronomy, Space Science and Astrophysics, Independent University, Bangladesh, Dhaka 1229, Bangladesh}
\email[show]{slahad@uwaterloo.ca}  

\author[orcid=0009-0006-6412-6245,sname='Reid']{Rashaad Reid} 
\affiliation{Waterloo Centre for Astrophysics, University of Waterloo, Waterloo, ON N2L 3G1, Canada}
\affiliation{Department of Physics and Astronomy, University of Waterloo, 200 University Avenue West, Waterloo, Ontario N2L 3G1, Canada}
\email{yreid@uwaterloo.ca}

\author[orcid=0000-0002-7805-2500,sname=Mpetha]{Charlie T. Mpetha}
\affiliation{NASA Goddard Space Flight Center, 8800 Greenbelt Rd, Greenbelt, MD 20771, USA}
\email{charlie.mpetha@nasa.gov}

\author[orcid=0000-0002-6639-4183, sname=Taylor]{James E. Taylor}
\affiliation{Waterloo Centre for Astrophysics, University of Waterloo, Waterloo, ON N2L 3G1, Canada}
\affiliation{Department of Physics and Astronomy, University of Waterloo, 200 University Avenue West, Waterloo, Ontario N2L 3G1, Canada}
\email{taylor@uwaterloo.ca}

\author[orcid=0000-0002-9814-3338, sname=Hildebrandt]{Hendrik Hildebrandt}
\affiliation{Ruhr University Bochum, Faculty of Physics and Astronomy, Astronomical Institute (AIRUB), German Centre for Cosmological Lensing, 44780 Bochum, Germany}
\email{hendrik@astro.ruhr-uni-bochum.de}

\author[orcid=0000-0002-1437-3786, sname=Hudson]{Michael J. Hudson}
\affiliation{Department of Physics and Astronomy, University of Waterloo, 200 University Avenue West, Waterloo, Ontario N2L 3G1, Canada}
\affiliation{Waterloo Centre for Astrophysics, University of Waterloo, Waterloo, ON N2L 3G1, Canada}
\affiliation{Perimeter Institute for Theoretical Physics, 31 Caroline St. North, Waterloo, ON N2L 2Y5, Canada}
\email{mike.hudson@uwaterloo.ca}

\author[orcid=0000-0001-6965-7789, sname=Chambers]{Kenneth C. Chambers}
\affiliation{Institute for Astronomy, University of Hawaii, 2680 Woodlawn Drive, Honolulu, HI 96822, USA}
\email{chambers@ifa.hawaii.edu}

\author[orcid=0000-0001-5486-2747, sname=de Boer]{Thomas de Boer}
\affiliation{Institute for Astronomy, University of Hawaii, 2680 Woodlawn Drive, Honolulu, HI 96822, USA}
\email{tdeboer@hawaii.edu}

\author[orcid=0009-0004-3655-4870, sname=Guerrini]{Sacha Guerrini}
\affiliation{Universit\'e Paris-Saclay, Universit\'e Paris Cité, CEA, CNRS, AIM, 91191, Gif-sur-Yvette, France}
\email{sacha.guerrini@cea.fr}

\author[orcid=0000-0002-5068-7918, sname=Guinot]{Axel Guinot}
\affiliation{Department of Physics, McWilliams Center for Cosmology, Carnegie Mellon University, Pittsburgh, PA 15213, USA}
\email{aguinot@andrew.cmu.edu}

\author[orcid=0000-0001-8221-8406, sname=Gwyn]{Stephen Gwyn}
\affiliation{National Research Council Herzberg Astronomy and Astrophysics, 5071 West Saanich Road, Victoria, B.C., V8Z6M7, Canada}
\email{Stephen.Gwyn@nrc-cnrc.gc.ca}

\author[orcid=0000-0001-9513-7138, sname=Kilbinger]{Martin Kilbinger}
\affiliation{Universit\'e Paris-Saclay, Universit\'e Paris Cité, CEA, CNRS, AIM, 91191, Gif-sur-Yvette, France}
\email{martin.kilbinger@cea.fr}

\author[orcid=0000-0002-2637-8728, sname=Van Waerbeke]{Ludovic Van Waerbeke}
\affiliation{Department of physics and astronomy, University of British Columbia, 6224 Agricultural road, V6T 1Z1 Vancouver, Canada}
\email{waerbeke@gmail.com}

\begin{abstract}

We investigate how the dynamical state of galaxy clusters influences their galaxy populations and mass distributions. Using photometrically selected clusters from the DESI Legacy Imaging Survey cross-matched with the UNIONS galaxy shear catalogue, we classify clusters as \textit{evolved} or \textit{evolving} based on their rest-frame $r$-band magnitude gaps and stellar mass ratios between the brightest cluster galaxies (BCGs) and bright satellites. We measure the stellar mass functions, weak-lensing profiles, and radial number density and red-fraction profiles of stacked clusters in both subsamples. Evolved clusters exhibit more concentrated lensing profiles, bimodal stellar mass functions dominated by massive BCGs, and a deficit of intermediate-mass satellites, while evolving clusters show flatter central lensing signals and an excess of massive satellites. Applying the same selection to IllustrisTNG clusters reproduces these trends and links the observed differences to distinct mass accretion histories. These results demonstrate the close link between cluster galaxy populations and the overall dynamical state of their underlying dark matter halo.

\end{abstract}


\section{Introduction}
 
 As the largest virialized structures in the Universe, galaxy clusters provide important tests of cosmology, dark matter, fundamental physics and galaxy evolution \citep[e.g.][]{Allen2011,Eckert2022,Vogt2024,Roche2024,Butt2025,Chen2025}. Clusters are detectable out to high redshift using a variety of methods, including X-ray searches for extended sources \citep[e.g.][]{Adami2018,Bulbul2024}, measurements of a Sunyaev-Zeldovich decrement at millimeter wavelengths \citep[e.g.][]{Planck2016b,Hilton2021,Bleem2024}, and clustering of galaxies in optical or infrared photometric and spectroscopic surveys \citep[e.g.][]{Rykoff2014,Radovich2017,Oguri2018,McClintock2019}. Current and forthcoming surveys using each of these detection methods are expected to produce catalogues of tens or hundreds of thousands of clusters \citep[e.g.][]{LSST2009,DESI2016,Sartoris2016, wen2024}, increasing previous sample sizes by an order of magnitude or more. Thus, we can expect the precision tests of physics and cosmology based on cluster properties to improve dramatically.

Essentially all methods of cluster detection depend sensitively on mass and/or redshift, and thus cluster properties are often considered in bins of mass and redshift, or even in aggregate over a whole sample. This is true of most studies of cluster abundance \citep[e.g.][]{Planck2016a,Pacaud2018,Ghirardini2024, Bocquet2024}, but also of mass-observable scaling relations \citep{Vikhlinin2009, Mantz2010, Bocquet2019}, stacked X-ray, SZ or lensing profiles \citep[e.g.][]{Giocoli2024,Popesso2024}, and studies of cluster galaxy populations \citep[e.g.][]{Zenteno2016,Valk2025}. However, clusters continue to grow through major mergers right down to the present epoch, and individual clusters should vary greatly in their accretion histories. Accretion history is known to influence observable physical properties and likely has systematic effects on scaling relations, impacting the robustness of cosmological inferences \citep{Rykoff2014, Herbonnet2020, Grandis2021}. Thus, as sample sizes increase and tests become more precise, cluster studies should attempt to control for this dependence more carefully. 

Beyond introducing systematics in mass estimates and scaling relations, cluster structural properties themselves also encode cosmological information. Morphological and dynamical state metrics — such as substructure fractions, concentration, and offsets between X-ray, lensing, or Brightest Cluster Galaxy (BCG) centres — depend on cosmology through its influence on growth rates \citep{Amoura2021, Lau2021,Amoura2024,Haggar2024,Mpetha2024}. Cluster formation history also influences their constituent galaxy populations. The quenching of satellites, the assembly of brightest cluster galaxies, and the buildup of intracluster light all depend on the accretion history of the halo \citep{DeLucia2007,Donnari2021,ahad2023,brough2024,kimmig2025}. Understanding cluster structure and its connection to growth history is therefore crucial not only for cosmology, but also for models of galaxy evolution across the protocluster–to–cluster transition \citep{Chiang2013,Overzier2016}.

Unlike mass or redshift, growth history is not directly measurable, nor can it be fully described by a single variable. Simulations do suggest, however, that some structural properties should be strongly correlated with specific aspects of the assembly history. In particular, the magnitude gap between the first and second-ranked galaxies, or other similar magnitude gaps or stellar mass ratios are found to be strongly correlated with the half mass accretion redshift $z_{50}$ \citep{kimmig2025}. Thus, by measuring these properties and selecting on them, we may be able to control for the effect of growth history on cluster properties.

Here, we apply this strategy for the first time to a very large (10$^4$ object) sample of clusters. Separating out subsamples of systems with the largest and smallest magnitude gaps and mass ratios, we show that not only do the overall optical properties differ significantly between the two subsamples, but so do the mean lensing profiles. Comparison with analogous measurements made on simulated clusters suggests that these observed differences are due to systematic differences in growth histories. It also suggests that mass bias, projection effects and a number of other systematics should be correlated with growth history, and thus with observable structural or dynamical parameters.

The outline of the paper is as follows. In Section \ref{sec:data} we describe the basic data and initial cluster catalogues used in this study. In Section \ref{sec:cl_sample}, we explain how these are combined to produce a final cluster catalogue, from which we select `evolved' (relaxed) and `evolving' (unrelaxed) subsets.
In Section \ref{sec:r_ur_properties}, we compare the optical properties and lensing profiles of these two subsets. We then compare the observed properties to predictions from cosmological simulations in Section \ref{sec:illustris}. We discuss our results in Section \ref{sec:discussion} and conclude in Section \ref{sec:conclusions}. 

A flat $\Lambda$CDM cosmology is assumed for any relevant calculations in this work, with $H_0$ = 70 kms$^{-1}$Mpc$^{-1}$, $\Omega_{\Lambda} = 0.7$, and $\Omega_\textrm{M} = 0.3$. 

\section{Data and initial catalogues}
\label{sec:data}

To robustly investigate the impact of cluster formation history on cluster structure and the properties of member galaxies, we require a large and homogeneous sample of galaxy clusters derived from observations. Wide-field, multi-wavelength surveys now provide extensive sky coverage and value-added cluster catalogues suitable for such analysis. For this work, we employ galaxy cluster catalogues constructed from photometric redshifts in the Dark Energy Spectroscopic Instrument (DESI) Legacy Imaging Surveys \citep[DLIS;][]{Dey2019}.

Our final sample is drawn from a cross-matched catalogue between two DLIS-based cluster catalogues, from \citet{wen2024} (hereafter \citetalias{wen2024}) and \citet{yang2021} (hereafter \citetalias{yang2021}), together with the weak-lensing shear measurements of the Ultraviolet Near-Infrared Optical Northern Survey \citep[UNIONS;][]{gwyn2025}. The individual surveys and catalogues are described in the following sections.

\subsection{Cluster catalogues based on DLIS data}
\label{sec:desi_ls_catalogues}

\subsubsection{Wen \& Han 2024 cluster catalogue}
\label{sec:wh24_sample}

The primary cluster catalogue we used is the one reported by \citetalias{wen2024}, containing 1.58 million galaxy clusters from the DLIS survey. Their cluster finding algorithm is based on \citet{Wen2021}, and has two main steps: (i) identifying brightest cluster galaxy (BCG) candidates from the DESI Legacy survey galaxy catalogue; and then (ii) detecting galaxy clusters by searching for overdensities of stellar mass around these potential BCGs within a redshift slice. Their BCG selection is based on the galaxy stellar mass, colour ($r-z_m$), and magnitude ($z_m, W_1$ bands). \citetalias{wen2024} primarily use photometric redshifts of galaxies to detect their BCG candidates and cluster members around them. These photometric redshifts are derived from two independent sources, (i) \citetalias{Dey2019} DR9 photometric redshifts using the $grzW1W2$ magnitudes and a random forest algorithm by \citet{Zhou2021}; and (ii) \citetalias{Dey2019} DR10 photometric redshifts using the $grizW1W2$ magnitudes based on a nearest neighbour algorithm described in \citet{Wen2021}. For a subset of these galaxies, which are predominantly the BCG candidates, they use spectroscopic redshifts, where available. We restricted our sample to clusters with a spectroscopically estimated redshift for the BCG, which is also reported as the redshift of the corresponding cluster.

We first choose the \citetalias{wen2024} clusters that are within the UNIONS $r-$band imaging coverage and have a spectroscopic redshift for the BCG. This selection reduces the cluster sample to about 110000. To reduce the inclusion of spurious cluster detections, we select only clusters with at least 10 member galaxies within their $R_{500c}$. We further limit the sample to clusters $z \leq 0.5$ to be able to apply reliable k-corrections to the cluster galaxy magnitudes in $r-$band, which is crucial for selecting clusters for the primary analysis in this work (discussed in detail in Section~\ref{sec:cl_sample}). The redshift limit also guarantees that all of the cluster BCGs have a $z-$ band magnitude $m_z < 21$, which ensures the best photometric redshift estimates from the \citetalias{wen2024} sample (figure 1 of \citetalias{wen2024}). The final cluster sample from this dataset, after imposing the above-mentioned constraints, is 30564. 

The cluster catalogue accompanying \citetalias{wen2024} includes estimates of various properties for the selected clusters, of which we use the positions, redshifts, cluster halo mass ($M_{500c}$), radius ($R_{500c}$), and number of member galaxies within $R_{500c}$ ($N_\mathrm{clmem}$). They also provide an augmented catalogue of cluster member galaxies, including their positions, stellar masses, magnitudes in $grizW1W2$ bands, and distances from the cluster centre. We obtain the positions, stellar mass, and $r-$band apparent magnitudes of the cluster members directly from this catalogue. 

\citetalias{wen2024} estimate the cluster halo mass from a total-stellar-mass-to-halo-mass relationship. First, they estimate the stellar mass within a radius $[E(z)^2]^{-1/3}$~Mpc around the BCG (where $E(z)^{2} = \rho_c(z)/\rho_c(0) =  \mathrm{\Omega}_{\Lambda} + \mathrm{\Omega}_m (1+z)^3$ captures the evolution of the critical density $\rho_c$ with redshift). They then estimate $R_{500c}$ through a scaling relation calibrated and reported in \citet{Wen2021}. Finally, they measure the stellar mass in cluster member galaxies within $R_{500c}$,  $m_{\star,500}$, and obtain the total mass $M_{500c}$ from the scaling relation $\mathrm{log} (M_{500c}/10^{14}\text{M}_{\odot}) = 0.96\ \mathrm{log}\ (m_{\star,500}/10^{10}\text{M}_{\odot}) - (1.82 \pm 0.02)$ reported in \citet{Wen2021}. They estimate the galaxy stellar mass using independent scaling relations between the $W1-$ and $z-$band luminosities and the stellar mass separately, and take the average of the two estimates. These scaling relations are calibrated to the COSMOS2015 catalogue \citep{laigle2016}. For galaxy cluster detection in \citetalias{wen2024}, they only use galaxies with a stellar mass of at least $10^{10}\ \mathrm{M}_{\odot}$.

Although the authors report high completeness across the halo mass range, the catalogue may still include contamination from spurious detections caused by interloper galaxies projected along the line of sight. To increase the purity of our cluster selection, we cross-match our sample to a cluster catalogue  constructed independently from the DLIS galaxy data, as described below. 

\subsubsection{Yang et al. 2021 cluster catalogue}
\label{sec:y21_sample}

The \citetalias{yang2021} group catalogue is constructed from the DLIS DR8 galaxy sample, augmented with spectroscopic redshift information from the DESI DR1 \citep{desidr2025} and other available surveys, where applicable. This catalogue extends the halo-based group finder originally developed by \citet{Yang2005,Yang2008} to simultaneously handle photometric and spectroscopic redshifts. The halo-based group finder connects galaxies that are likely to be within the same dark matter halo using an iterative abundance-matching approach. Each galaxy is first treated as a tentative group centre, and the algorithm iteratively refines group membership and halo properties based on luminosity, redshift, and spatial proximity. Their photometric redshifts are estimated using the $grzW1W2$ magnitudes from DLIS DR8 with the Photometric Redshifts for the Legacy Surveys machine-learning algorithm of \citet{Zhou2021}. They assign cluster halo masses ($M_{180m}$) through a luminosity-based abundance matching, which is calibrated on the mock catalogue from the ELUCID simulations \citep{Wang2016}. We apply a mass mean conversion function $M_{500c} = 0.55\ M_{180}$ based on \citet{hu2003} to obtain $M_{500c}$ from their reported halo mass. They define the cluster centre as the luminosity-weighted centre of the cluster member galaxies.

For this work, we use Y21 clusters that lie within the UNIONS $r$-band imaging footprint and have at least 10 detected member galaxies. We further restrict ourselves to systems with $z\leq0.5$ and $M_{500c}\geq 10^{13.0} h^{-1}\mathrm{M}_{\odot}$ to select a sample comparable to our WH24 sample. After applying all of our selection criteria to the Y21 catalogue, we obtain a subsample of 83806 potential matches. For each cluster, the \citetalias{yang2021} catalogue provides the central coordinates, redshift, halo mass, total group luminosity, and richness. We use this sample to validate the cluster detection and obtain the halo mass estimation of the WH24 clusters, but do not use the \citetalias{yang2021} cluster member catalogue, as it lacks stellar mass estimates. Further justification for using the halo mass estimation from this catalogue for our final sample is provided in Sec.~\ref{sec:mass_comparison_wh24_y21}.

Although the Y21 algorithm demonstrates high completeness and purity, its reliance on photometric redshifts can still lead to projection-induced contamination. Nevertheless, its physically motivated halo-based approach and augmented spectroscopic redshifts from DESI DR1 make it an excellent complementary dataset to the BCG-centred \citetalias{wen2024} catalogue.

\subsection{UNIONS survey data}
\label{sec:unions_shear_cat}

The Ultraviolet Near Infrared Optical Northern Survey \citep[UNIONS;][]{gwyn2025} is a collaboration of wide-field imaging surveys of the northern hemisphere. UNIONS consists of the Canada-France Imaging Survey (CFIS), conducted at the 3.6-meter CFHT on Maunakea, members of the Pan-STARRS team, and the Wide Imaging with Subaru HyperSuprime-Cam of the Euclid Sky (WISHES) team. CFHT/CFIS is obtaining deep $u-$ and $r$ bands; Pan-STARRS is obtaining deep $i$ and moderate-deep $z$ band imaging, and Subaru is obtaining deep $z-$band imaging through WISHES and $g-$band imaging through the Waterloo-Hawaii IfA $g-$band Survey (WHIGS). These independent efforts are directed, in part, to securing optical imaging to complement the \emph{Euclid} space mission \citep{euclid2025}, although UNIONS is a separate collaboration aimed at maximizing the science return of these large and deep surveys of the northern skies.

In this work, we use the weak-lensing galaxy shear catalogue from the UNIONS collaboration. We also only use clusters within the UNIONS $r-$band footprint to have consistent coverage for follow-up projects. The shape catalogue was created using the publicly available \texttt{ShapePipe} software \citep{Farrens2022,guinot2022} on $r$-band images taken by MegaCam on CFHT. We use version 1.4.5 of the UNIONS ShapePipe catalogue, which contains $85\,$million sources over $\sim\!3,\!200\,$deg$^2$, giving a raw number density of $7.4\,$arcmin$^{-2}$.

\section{Cluster sample selection}
\label{sec:cl_sample}

To improve the purity of our cluster sample, we cross-match the initially selected cluster samples from \citetalias{wen2024} and \citetalias{yang2021} described in Sec.~\ref{sec:wh24_sample} and~\ref{sec:y21_sample}. As these two catalogues define the cluster centres in different ways, there can be positional mismatches of up to a few hundred kpc between them, particularly for clusters that have gone through a merger in the past few Gyrs and are still dynamically disturbed. Therefore, to cross-match these catalogues, we use a projected separation tolerance of 300~physical kpc between the cluster centres at the cluster redshift, and a redshift offset tolerance of $\Delta z/(1+z)\leq0.02$, which is equivalent to $\delta v \sim 6000$\,km/s. (The redshift tolerance corresponds to an offset in velocity several times the velocity dispersion of a typical cluster, but reflects the fact that both catalogues are primarily based on photometric redshifts, and include potential cluster member galaxies selected with even higher redshift tolerance to match the photometric redshift uncertainties.) We picked the closest match in both physical and redshift space within these tolerances. After the matching, we find 16873 uniquely matched clusters in the final sample. 

\subsection{Selection of evolved and evolving systems}

Recent studies have shown that the dynamical state of galaxy clusters can be traced by many different indicators, which are connected to different aspects of the growth history  \citep[e.g.][]{cui2017,kimmig2023,Haggar2024,kimmig2025,veliz-astudillo2025}. 
In this work, we are particularly interested in the overall assembly history, and whether the cluster is still `evolving' significantly through recent accretion, or is it already well `evolved', having accreted most of its mass long ago. In simulations, one of the simplest indicators of the mass accretion history is the cluster formation redshift or ($z_{form}$ or $z_x$), which refers to the redshift by which $x\%$ of the cluster's final mass was assembled into a single halo; $x$ can be 50, 75, 90, or any other number depending on the chosen definition  \citep[e.g.][]{Lacey1993,Wong2012}. This parameter cannot be measured directly in observations; however, using a large sample of galaxy clusters taken from several large volume hydrodynamical simulations, \citet{kimmig2025} have shown that a number of observable indicators of a cluster's dynamical state show good correlation with $z_{50}$. These indicators include: i) the stellar mass ratio between the BCG and the second most massive galaxy within $R_{200c}$ of the cluster, $M_{\star, 12} = M_{\star, \textrm{BCG}}/M_{\star,2}$ \citep{tremaine1977,Loh2006,ragagnin2019}; ii) the stellar mass ratio between the BCG and the fourth most massive galaxy in the cluster, $M_{\star, 14}$ \citep{goldenmarx2018}; iii) the normalized offset $\Delta_r/R_{200c}$ between the centre of stellar mass of the cluster within $R_{200c}$ and the position of the BCG, also known as the center shift \citep{mann2012,biffi2016,contreras-santos2022}, and analogous to the theoretically defined offset parameter of \citet{Maccio2007}; and iv) the central stellar gravitational potential depth of the BCG, defined as the ratio between the BCG stellar mass $M_{\star, \textrm{BCG}}$ and the stellar half-mass radius $R_{\star,\ 1/2,\ \textrm{BCG}}$ for the BCG within $0.1 \times R_{200c}$ \citep{bluck2023}. The stellar mass ratio in these simulations is taken to be analogous to the $r-$band  magnitude gap (or flux ratio) between the BCG and the second brightest galaxy in the cluster. This ratio is also known as the fossil-ness parameter, as defined by \citet{voevodkin2010}. We use the absolute $r-$band magnitude gap between BCG and the second brightest galaxy ($M_{r,12}$), and the stellar mass ratio between the BCG and the fourth most massive galaxy ($M_{\star, 14}$) to select our evolved and evolving samples from the cross-matched clusters because these observable features can be robustly measured even when a cluster identifier fails to detect some fainter cluster members. In simulations, both parameters are positively correlated with the cluster formation history, such that, higher values of $M_{r,12}$ and $M_{\star,14}$ indicate more evolved and relaxed clusters, while lower values indicate more disturbed and rapidly accreting systems \citep[][]{voevodkin2010,ragagnin2019,kimmig2025}. 

We use the apparent $r-$band magnitudes, stellar masses, and photometric redshifts of the cluster members provided by the \citetalias{wen2024} catalogue to calculate $M_{r,12}$ and $M_{\star,14}$ for the initially selected 16873 clusters. To measure $M_{r,12}$, we apply k-correction on all the cluster member galaxies and estimate their absolute $r-$band magnitudes using the function: $M_x = m_x - \mu(z) - kcorr_x$, where $x$ corresponds to the optical bands $g$ and $r$, $M_x$ is the absolute magnitude, $m_x$ is the apparent magnitude, and $\mu(z)$ is the distance modulus at the galaxy's redshift. 

The k-correction for each band ($kcorr_x$) is measured from the corresponding apparent magnitudes and their $g-r$ colours using the method from \citet{chilingarian2010}. Finally, for each cluster, the satellite galaxies were ranked by their $r-$band absolute magnitudes, and the magnitude gap $M_{r,12}$ was measured from the absolute magnitudes of the BCG and the brightest satellite.

The stellar mass ratio is calculated by ranking the satellite galaxies from each cluster according to their stellar mass $\log (M_{\star})$, and then subtracting the $\log (M_{\star})$ of the third most massive satellite (hence fourth massive cluster member considering BCG as the rank-1 galaxy) from the $\log (M_{\star})$ of the BCG. 

\begin{figure}
    \centering
    \includegraphics[width=\columnwidth]{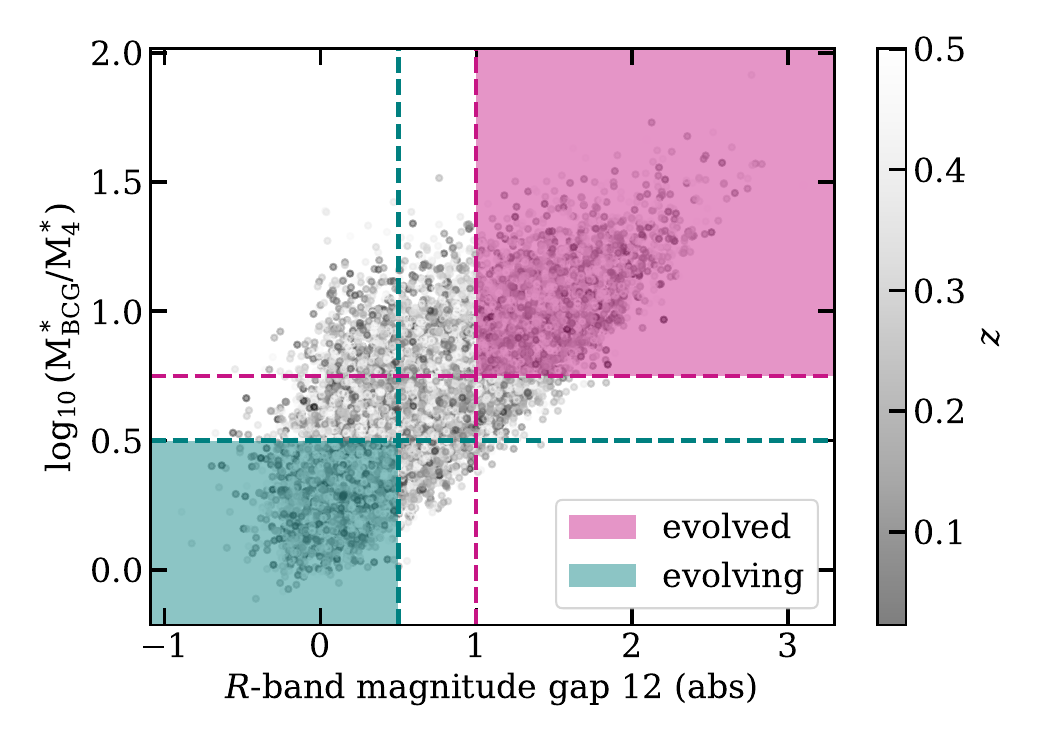}
    \caption{Selection of the evolved and evolving subsamples. The $x$-axis shows the absolute $r-$band magnitude gap between the BCG and the brightest satellite galaxy and the $y$-axis shows the log of the stellar mass ratio of the BCG to the 3rd most massive satellite of each galaxy cluster. The dashed lines and shaded regions indicate our selection regions for the evolved and evolving samples.}
    \label{fig:cl_selection_criteria}%
 \end{figure}
 
The distribution of stellar mass ratio versus magnitude gap for our cross-matched cluster sample is shown in Fig.~\ref{fig:cl_selection_criteria}. The colourbar on the right shows the sample redshift distribution. 
As expected, these two parameters have a positive correlation with some scatter. Our goal is to select subsamples of the most evolved and most actively evolving clusters from the sample. Therefore, we select the opposite ends of the $M_{\star,14}$ vs.~$M_{r,12}$ distributions: clusters with the highest values for both of these parameters are selected as the `evolved' subsample (pink shaded region in Fig.~\ref{fig:cl_selection_criteria}) and clusters with the lowest values for both of these parameters are selected as the `evolving' subsample (green shaded region in Fig.~\ref{fig:cl_selection_criteria}). The selection thresholds (shown by the dashed lines) were arbitrarily chosen to include about 20\% of the entire sample in each subsample \footnote{We tested varying this selection to include between 15 and 25~\% of the whole sample in each of the evolved and evolving subsamples. All the findings from this work remain consistent across this variation. }. After this selection, we have 3967 evolved and 3850 evolving clusters. An independent test of our cluster dynamical-state selection, by comparing it with X-ray-identified clusters from eROSITA \citep{Bulbul2024}, is provided in Appendix~\ref{app:erosita_comparison}. 

\begin{figure}
    \centering
    \includegraphics[width=\columnwidth]{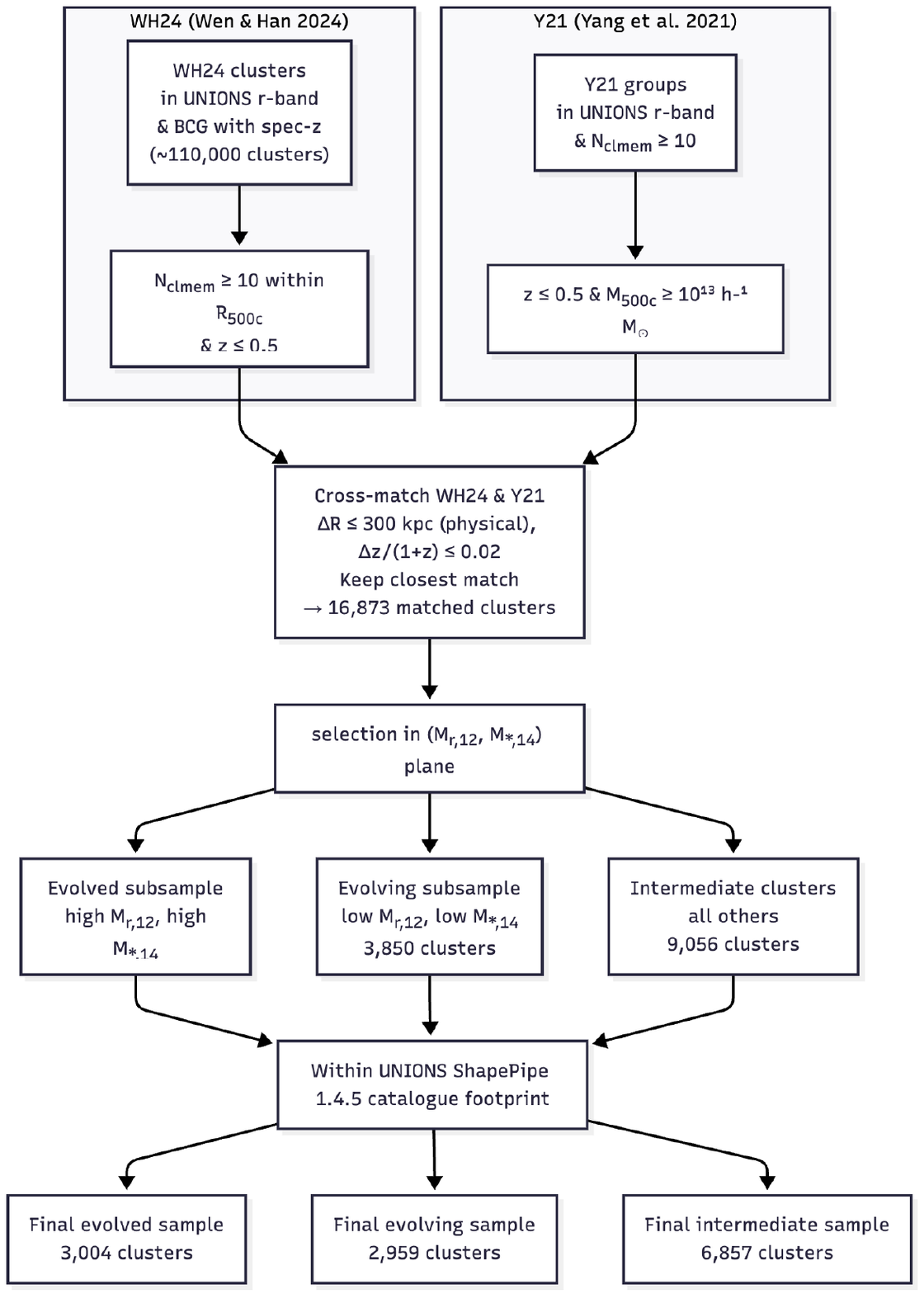}
    \caption{Flowchart showing our sample selection process combining multiple catalogues.}
    \label{fig:sample_selection_flowchart}%
 \end{figure}

Finally, we check whether the clusters are within the footprint of the UNIONS galaxy shear catalogue to obtain weak-lensing mass measurements and keep the samples consistent across all of our analyses. After this cross-matching, our final sample consists of 3004 evolved and 2959 evolving clusters, which will be referred to as ``the observed sample'' in the following sections. We also consider all the other clusters from Fig.~\ref{fig:cl_selection_criteria} that are not classified as evolved or evolving as `intermediate' clusters. We have a total of 9056 and 6857 intermediate clusters before and after matching to the UNIONS shear catalogue. The complete sample selection process is shown in Fig.~\ref{fig:sample_selection_flowchart}.

\begin{figure*}
    \centering
    \includegraphics[width=1.8\columnwidth]{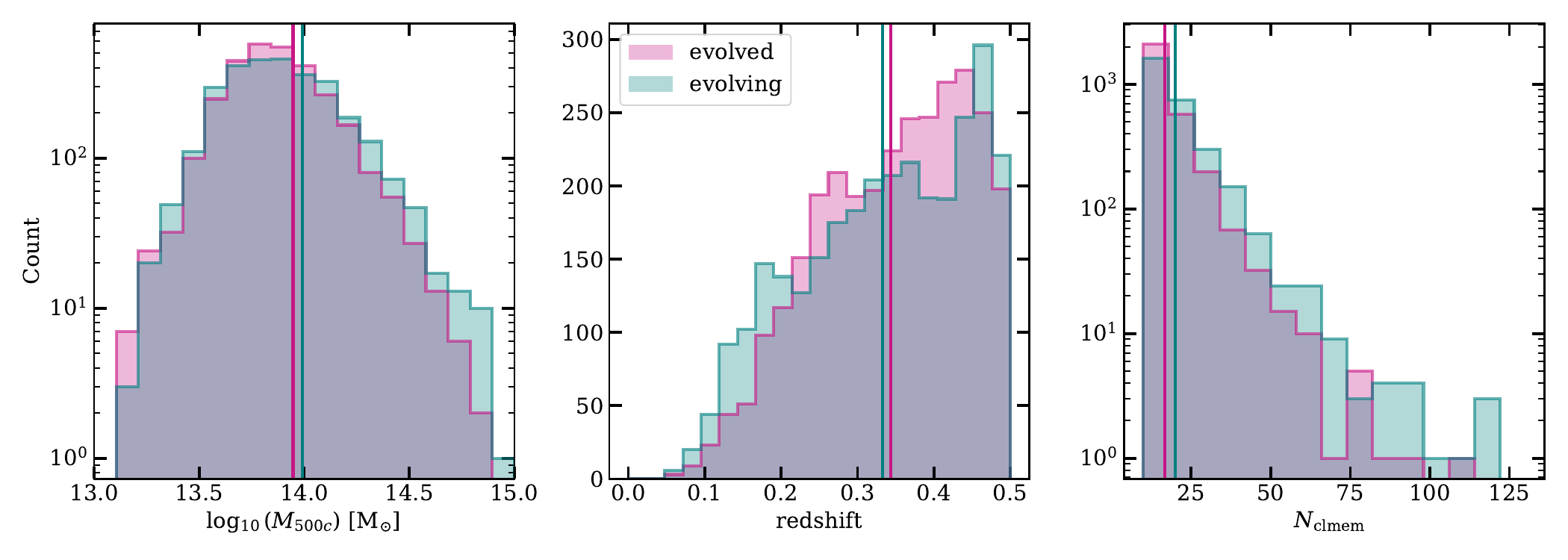}
    \caption{Distribution of cluster properties in the evolved and evolving samples. From left to right: Halo mass (average value 13.95 and 13.99 for evolved and evolving samples), redshift (average value 0.34 and 0.33 for evolved and evolving samples), and number of member galaxies within R$_{500c}$ (average value 16.7 and 20.1 for evolved and evolving samples), respectively. The average values of each sample are shown by vertical lines of the corresponding colours in each panel.}
    \label{fig:cl_properties_r_ur}%
 \end{figure*}

Figure~\ref{fig:cl_properties_r_ur} shows the distribution of halo mass ($M_{500c}$, left), redshift (middle), and number of member galaxies within $R_{500c}$ ($N_{\mathrm{clmem}}$, right) of our evolved (pink) and evolving (green) samples. The halo mass distribution indicates that they span the range from rich groups to massive clusters, with low-mass clusters with $M_{500c} \approx 10^{14}\mathrm{M}_{\odot}$ dominating the counts. The median redshift of both subsamples is about 0.3, and most of the clusters have $N_{\mathrm{clmem}}$ less than 50. There is no significant difference in any of these fundamental cluster properties between the evolved and evolving samples, demonstrating that we do not expect any bias in the samples driven by these properties. The distribution of these clusters in the UNIONS footprint (not shown here) also indicates no significant difference between the locations of the two samples. 

\subsection{Finding large-scale-structure around the clusters}
\label{sec:lss_galaxy_search}

The cluster member catalogue provided with \citetalias{wen2024} only contains members within $R_{500c}$ of the cluster centre, which is taken to be the BCG location. However, galaxy cluster properties, especially their mass accretion history, depend on their large-scale environments as well \citep[e.g.][]{wetzel2012,bahe2013,bahe2017hydrangea}. To obtain more insight into the large-scale environment around our cluster samples, we therefore select galaxies within $1-10\times R_{500c}$ around our evolved and evolving cluster sample from the DLIS DR9-based galaxy photometric redshift catalogue of \citet{Zhou2021}. This galaxy catalogue was also used in the \citetalias{wen2024} cluster finding process; details of the photometric redshift measurement are given in Sec .~\ref{sec:wh24_sample}. We use the same photometric redshift slice around the BCG as \citetalias{wen2024} in this search for consistency. We find 1008202 and 1032125 galaxies within a circular disk with radial range $1-10\times R_{500c}$ around our evolved and evolving cluster samples, respectively. We discuss the impact of the cluster's large-scale environment on its weak-lensing profile and stellar mass function in Appendix~\ref{app:lss_impact}.

\section{Properties of evolved and evolving clusters}
\label{sec:r_ur_properties}

Galaxy clusters of similar halo mass and redshift are often stacked together to study ensemble properties such as the stellar mass function and weak-lensing mass profile. However, differences in their formation histories and dynamical states are expected to introduce significant scatter in these relations. In this section, we compare the stellar mass functions, the fraction of the stellar mass contained in the BCG, and the weak-lensing mass profiles of the `evolving' and `evolved' subsamples. Together, these diagnostics provide complementary insights into how baryonic and dark matter components co-evolve during cluster assembly.

\subsection{Stellar mass functions}
\label{sec:r_ur_smf}

One of the most fundamental observable properties of galaxies and clusters is their total stellar mass and the number distribution of galaxies in different stellar mass bins -- the stellar mass function (SMF). The SMF is the product of star formation, galaxy mergers, and stripping, so its shape and temporal evolution provide an observationally accessible tracer of galaxy evolution, and of the role of environment in this process. Thus, the SMF is a critical measurement against which galaxy formation models are tested \citep[e.g. ][]{ahad2021}.

\begin{figure*}
    \centering
    \includegraphics[width=\textwidth]{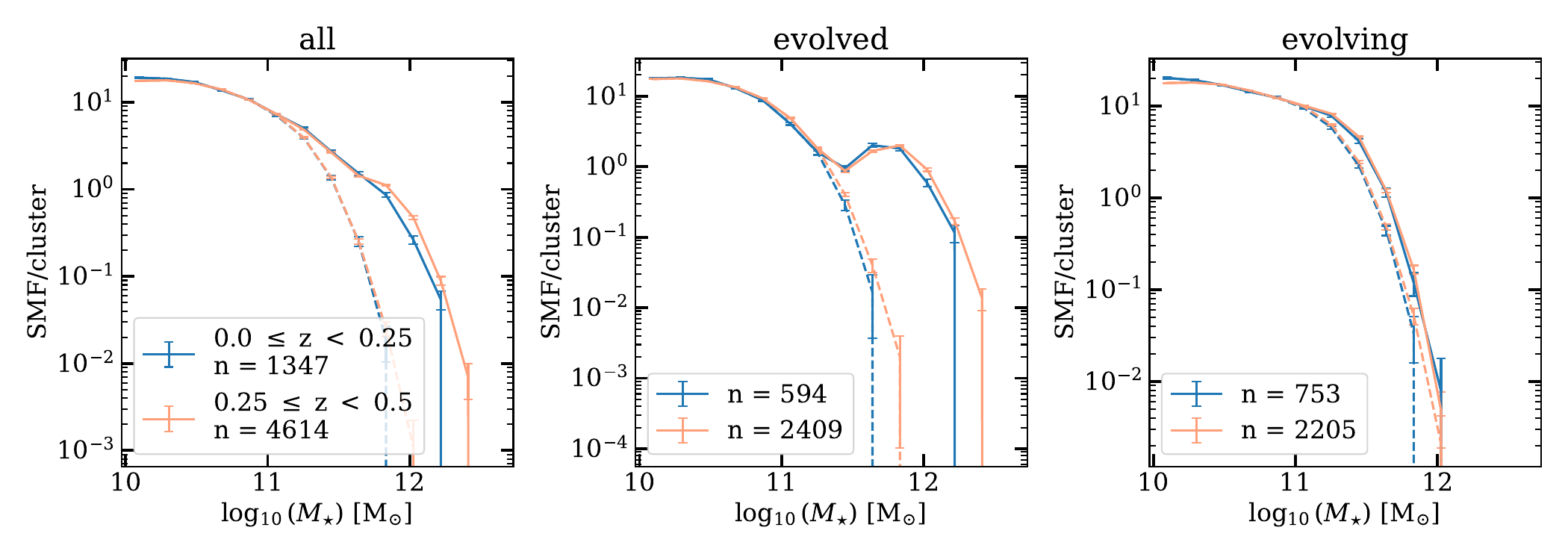}
    \caption{Stellar mass functions (SMFs) of cluster galaxies in two redshift bins ( $0.0 \leq z < 0.25$, blue; $0.25 \leq z < 0.5$, orange) for all (evolved+evolving) clusters (left), evolved clusters (middle), and evolving clusters (right). Solid lines include all members within $R_{500c}$, while dashed lines show SMFs of satellites only, excluding the BCG. Evolved systems display a bimodal SMF, with the high-mass peak corresponding to the BCGs reaching $\log_{10} (\mstar/\msun) \approx 12.5$, whereas BCGs in the evolving sample extend only to $\log_{10} (\mstar/\msun) \approx 12.0$.}
    \label{fig:smf_r_ur_sat}%
 \end{figure*}

To study any potential evolution of the SMF in our two subsamples, we first divide them into two redshift bins:  $0.0 \leq z < 0.25$ and $0.25 \leq z < 0.5$. For each of these redshift bins, we construct the SMF for the evolved and evolving subsamples separately, as well as measuring it for the combined (evolved and evolving) sample. 
Figure~\ref{fig:smf_r_ur_sat} shows the SMFs in the two redshift intervals for the full (left), evolved (middle), and evolving (right) cluster samples, respectively. The solid lines represent all galaxies within $R_{500c}$, while the dashed lines show only satellite members. The error bars are calculated from 100~bootstrap resamplings from the parent galaxy sample for each cluster subset, and show the $1$-$\sigma$ distribution of the measurements. The overall SMFs exhibit the familiar Schechter function-like shape \citep{schechter1976}, with a sharp decline at the high-mass end and almost no redshift evolution, consistent with passive stellar growth.

\begin{figure}
    \centering
    \includegraphics[width=\columnwidth]{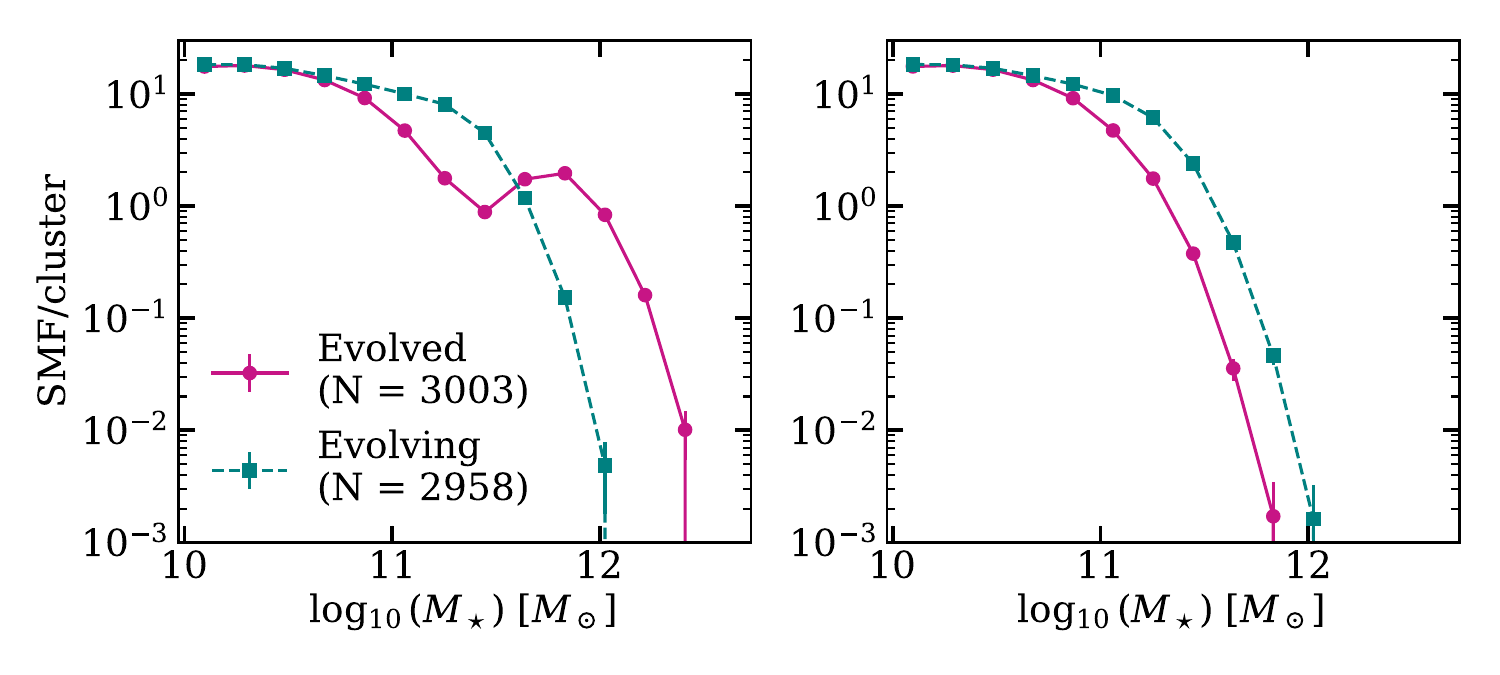}
    \caption{Stellar mass functions (SMFs) of the evolved (magenta) and evolving (teal) clusters across all redshifts. The left panel shows SMF with all cluster members, and the right panel shows satellites only. The SMF with all cluster members extend to larger stellar mass for the evolved sample, whereas the trend is opposite for the satellites. Either way, the SMFs are different for the evolved and evolving subsamples.}
    \label{fig:smf_allz}%
 \end{figure}

The shape and normalisation of the SMFs of the combined sample are consistent with the literature on cluster SMFs \citep[e.g.][]{vulcani2011,vdb2018}.
However, a clear difference emerges between the SMFs of the evolved and evolving systems. The evolved clusters display a bimodal SMF, where the more massive peak at $\log_{10} (\mstar/\msun) \approx 12.0$ represents the BCG population, separated from the satellite component by a pronounced magnitude gap. In contrast, the evolving clusters show a smoother transition between the BCG and satellite populations, and a peak of distribution at $\log_{10} (\mstar/\msun) \approx 11.4$. The bimodal SMF of the evolved systems and the smoother SMF of the evolving systems can be attributed to the selection process based on the magnitude gap between the BCG and the second brightest galaxy. However, the stellar mass ratio between BCG and the fourth massive galaxy does not directly predict a bimodal SMF for the evolved systems.

There are two other key differences between the evolved and evolving samples, which are not directly predictable from the selection functions. Firstly, the SMFs of evolved systems are extended out to $\log_{10} (\mstar/\msun) \approx 12.5$, whereas the BCGs of the evolving systems only range up to $\log_{10} (\mstar/\msun) \approx 12.0$. Secondly, the evolving systems also host a noticeable excess of intermediate-mass satellites ($\log_{10} (\mstar/\msun) \sim 10.8 - 11.3$) compared to the evolved systems, suggesting that the BCGs in dynamically evolved clusters may have grown at the expense of such massive satellites, through mergers and dynamical friction. This is better highlighted in Fig.~\ref{fig:smf_allz}, where the SMFs of the evolved and evolving samples are shown on the same panels for all cluster members, including BCG (left), and for satellites only (right). This behaviour supports the interpretation that the magnitude-gap–based classification effectively traces the hierarchical assembly stage of the cluster. We explore this further in the following section.

\subsection{BCG growth in evolved and evolving systems}
\label{sec:r_ur_bcg_growth}

\begin{figure*}
    \centering
    \includegraphics[width=2\columnwidth]{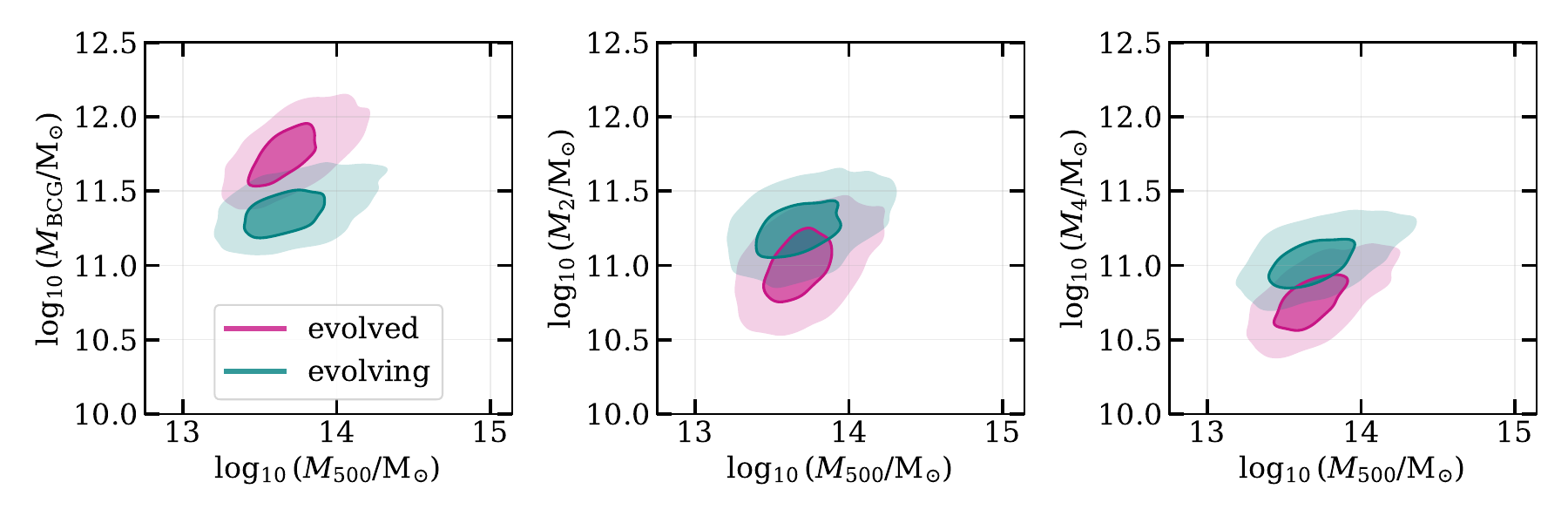}
    \caption{The 1- and 2-$\sigma$ distribution of the stellar mass of different ranked galaxies vs.~the total mass $M_{500}$ of the evolved and evolving clusters. From left to right, the y-axes show the stellar mass of the BCG, the second brightest cluster galaxy, and the fourth massive cluster galaxy within $R_{500}$.}
    \label{fig:BCG_M2_M4_mass_vs_halo_mass}
\end{figure*}

From Fig.~\ref{fig:smf_r_ur_sat}, evolved and evolving clusters show a significant difference in their BCG mass distributions, but these samples include systems with a range of halo masses.  
To clarify the relation between the top-ranked galaxy masses and halo mass, in Fig.~\ref{fig:BCG_M2_M4_mass_vs_halo_mass} we plot the distribution of the stellar mass of the BCG (left panel), the second most massive galaxy (middle panel), and the fourth most massive galaxy (right panel) vs.~the estimated halo mass of the cluster, in the evolved (pink) and evolving (teal) samples. In each of the panels, the inner darker and outer lighter shaded contours indicate the $1$- and $2$-$\sigma$ distribution of the corresponding cluster samples.

We find that, at fixed halo mass, evolved clusters host systematically more massive BCGs on average than evolving systems. This suggests that part of the observed scatter in the total-stellar-mass–halo-mass relation \citep[see also][]{Huang2020} in clusters may be driven by the magnitude gap; incorporating the magnitude gap could therefore reduce the scatter in the BCG-mass–halo-mass plane. Interestingly, the middle and right panels of Fig.~\ref{fig:BCG_M2_M4_mass_vs_halo_mass} show that the distribution of the second and fourth most massive galaxies in these systems follow an opposite trend, i.e., second and fourth-ranked galaxies in evolved clusters are, in general, less massive than those in evolving clusters. This trend again suggests a potential mass transfer from massive satellites to the BCGs in evolved systems. We need stronger evidence, however, to strengthen the connection between our observational selection method and the underlying state of the cluster. Thus, in the next section, we will consider the mean mass profiles of the two samples, as measured by gravitational lensing.

\begin{figure}
    \centering
    \includegraphics[width=0.95\columnwidth]{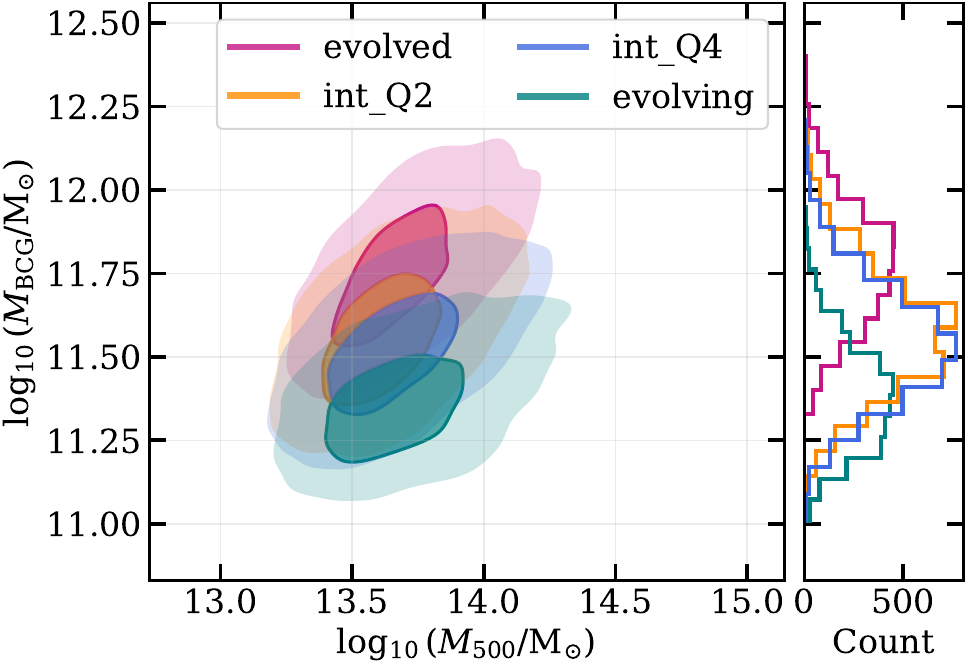}
    \caption{The 1- and 2-$\sigma$ distribution of the BCG stellar mass vs.~the $M_{500}$ of the different cluster sub-samples defined in the text.}
    \label{fig:BCG_mass_vs_halo_mass}
\end{figure}

The BCG-to-halo-mass distribution of the evolved and evolving systems naturally raises one more question: what happens to the clusters that are in between these two samples in Fig.~\ref{fig:cl_selection_criteria}? To determine this, along with the evolved and evolving subsamples, we divide the remaining clusters into two further subsamples. Clusters in the upper left corner of Fig.~\ref{fig:cl_selection_criteria} with a small magnitude gap and a large stellar mass ratio are defined as `intermediate clusters in quadrant 2' or `int\_Q2', and clusters in the lower right corner of Fig.~\ref{fig:cl_selection_criteria} with a large magnitude gap and small stellar mass ratio are defined as `intermediate clusters in quadrant 4' or `int\_Q4' samples. 

Figure~\ref{fig:BCG_mass_vs_halo_mass} shows the BCG stellar mass vs.~cluster halo mass distribution of the evolved, evolving, and intermediate samples. Similar to Fig.~\ref{fig:BCG_M2_M4_mass_vs_halo_mass}, the darker and lighter shaded regions here show the $1$- and $2$-$\sigma$ distribution of the corresponding cluster samples. The histograms on the right show the BCG stellar mass distribution of the same samples. As expected, the BCG vs.~halo mass distribution of the intermediate samples lies in between the evolved and evolving samples. The intermediate sample in $Q2$ has a slightly higher median BCG mass compared to the sample in $Q4$. The halo mass distribution of all four subsamples is similar, again hinting that the scatter in the stellar-to-halo-mass distribution for BCGs in cluster halos can be explained via the magnitude gaps in these systems. 

\subsection{Weak lensing profiles}
\label{sec:r_ur_weak_lensing}

We employ the current UNIONS weak lensing shape catalogue, ShapePipe v.1.4.5, to measure a stacked cluster lensing profile for each sample. The lensing profile can provide an independent verification of the dynamical state split, and give further information on the properties of these samples. 

The gravitational potential of a foreground galaxy cluster causes a shape distortion on the image of background galaxies. By measuring the correlated shape distortion of galaxies at a projected radius $R$ from the cluster centre, the surface mass density at $R$ can be inferred. Large numbers of galaxies are needed to overcome shot noise---stacking many clusters can provide a clean, significant measurement of their composite 2D mass profile. This projected mass density is expressed through the {\it excess surface mass density} $\Delta\Sigma$, 
\begin{equation}
        \Delta \Sigma(R) = \overline{\Sigma}(<R)-\Sigma(R) \, , \\
        \label{eq:dSigma}
\end{equation}
which is the difference between the average enclosed surface density at the projected radius $R$,
\begin{equation}
        \overline{\Sigma}(<R) = \frac{2}{R^2} \int_0^R R' \Sigma(R') \, dR' \, , \\
\end{equation}
and the surface density at $R$,
\begin{equation}
        \Sigma(R) = 2\int_R^\infty \frac{\rho(r)r}{\sqrt{r^2 - R^2}}\, dr \, ,
\end{equation} 
where the density profile of the stacked cluster at the 3D radius $r$ is given by $\rho(r)$, and is assumed to be spherically symmetric.

The $\Delta\Sigma$ profile has a simple theoretical link to the tangential shear, $\gamma_{\rm t}$, of background galaxies:
\begin{equation}
    \Delta\Sigma(R) = \gamma_{\rm t}(R) \Sigma_{\rm crit} \, ,
\end{equation}
where the critical surface mass density, $\Sigma_{\rm crit}$, between a lens `l' and source `s' with comoving distances $\chi$ is,
\begin{equation}
    \Sigma_{\rm crit, l-s} = \frac{c^2}{4\pi G}\frac{\chi_{\rm s}}{\chi_{\rm l}(\chi_{\rm s}-\chi_{\rm l})(1+z_{\rm l})} \, .
\end{equation}

A $\Delta\Sigma$ profile requires knowledge of the shear and the source and lens redshifts. However, we do not directly measure the shear, only the shapes of galaxies, $e_{\rm t}$. It is assumed there is no preferred galaxy shape, such that a large number of shapes can be averaged to obtain $\langle e_{\rm t} \rangle = g_{\rm t}$, where $g_{\rm t} = \gamma_{\rm t}/(1-\kappa)$ is the reduced shear. We make a further approximation that $g_{\rm t} \approx \gamma_{\rm t}$, given $\kappa$ is small over the range of radii of most interest. 

The calculation of $\Sigma_{\rm crit}$ is complicated by the lack of source redshifts in the ShapePipe v1.4.5 catalogue. Instead of computing $\Sigma_{\rm crit}$ for each lens-source pair, an effective survey average is found,
\begin{equation}
    \langle \Sigma_{\rm crit, l}^{-1}\rangle = \int \Sigma_{\rm crit, l-s}^{-1} n(z_{\rm s}) \, dz_{\rm s} \, . \label{eq:sigmacrit}
\end{equation}
The effective source redshift distribution used here, $n(z_{\rm s})$, has been constructed for the ShapePipe catalogue using the Self-Organising Maps \citep{Wright2020} method on a subset of galaxies matched to existing spectroscopic surveys. More detail can be found in Appendix~A of \cite{Li2024}. 

The observed $\Delta\Sigma$ profile, including appropriate weights for sources based on their shape noise (each lens is given equal weight), is
\begin{equation}
\Delta\Sigma(R) = \frac{\sum_{\rm l-s}w_{\rm s} e_{\rm t} \langle \Sigma_{\rm crit, l}^{-1}\rangle}{\sum_{\rm l-s}w_{\rm s}\langle \Sigma_{\rm crit, l}^{-1}\rangle^2} \, . \label{eq:esd}
\end{equation}
Two modifications to this raw profile are included. First, a set of random positions in the survey footprint $20\times$ larger than the number of clusters is chosen, and the lensing signal around these randoms is calculated, by assigning a random redshift drawn from the cluster $n(z)$ to each point. Subtracting the signal around randoms from the signal around lenses has been shown to mitigate the impact of systematics \citep{Mandelbaum2005}, and reduce the variance of the signal \citep{Singh2017}. Second, a boost factor correction is required to prevent the over-abundance of cluster members, which contain no lensing information, diluting the signal: 
\begin{equation}
    {\rm boost}(R) = \frac{\sum_{\rm l-s}w_{\rm s}\langle \Sigma_{\rm crit, l}^{-1}\rangle^2}{\sum_{\rm r-s}w_{\rm s}\langle \Sigma_{\rm crit, l}^{-1}\rangle^2} \, .
    \label{eq:boost}
\end{equation}
In each radial bin, the ratio between the abundance of galaxies for lens-source pairs and random-source pairs is used to estimate the over-abundance caused by cluster members, and this provides a correction factor to apply to the signal.

\begin{figure}
    \centering
    \includegraphics[width=1\columnwidth]{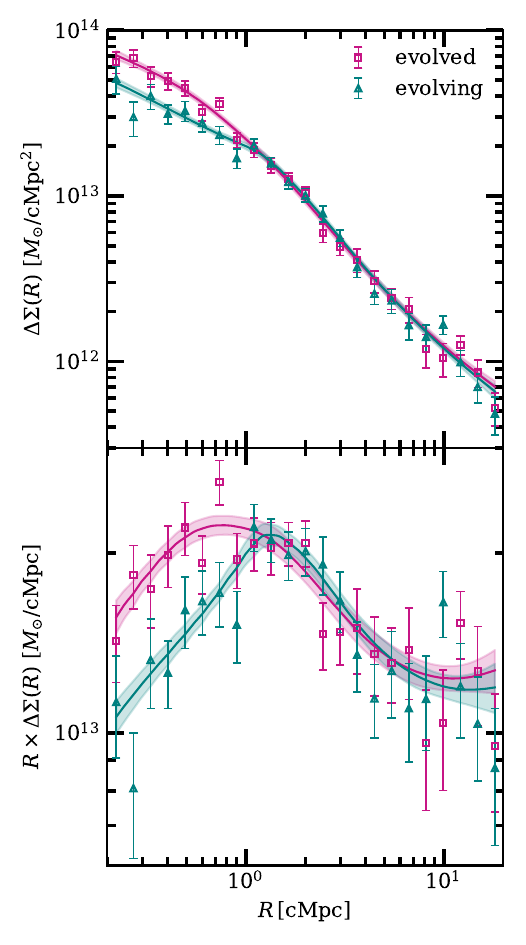}
    \caption{Weak lensing $\Delta\Sigma$ profiles of the evolved and evolving subsamples, all redshifts combined. Beyond 1~Mpc, their profiles are similar in amplitude, though with slightly different shapes. Within 1~Mpc, however, the evolved systems show a significantly stronger lensing signal. These profiles are calculated using the full evolved and evolving cluster samples over all redshifts.}
    \label{fig:lensing_profiles}%
 \end{figure}

A modified version of the \textsc{dsigma} package \citep{dsigma}, allowing the calculation of the cross-component of the shear $\gamma_{\times}$ is used to measure lensing profiles. We use 23 bins logarithmically spaced over the radial range $0.2 < R / {\rm cMpc} < 20$ (where cMpc denotes comoving present-day Mpc). The $\Delta\Sigma$ profiles are seen in Figure~\ref{fig:lensing_profiles}. There are significant differences between the mass profiles of the evolved and evolving samples in the inner region (note the cluster radius $r_{500{\rm c}} \simeq 0.7\,$Mpc for these samples). This difference is not a mass effect, as the mean mass of the two samples is very similar. We also measure the cross-component of the shear, and the signal around the random locations, verifying that they are both statistically consistent with zero.

As further verification, we fit these lensing profiles with a 3D density profile model \citep{Diemer2022}. The best-fit lines and $1\sigma$ bands are overlaid in Figure~\ref{fig:lensing_profiles}. The fitting procedure is outlined in detail in \cite{Mpetha2025}, and summarised in Appendix \ref{app:WL_fits}. Briefly, the density profile model includes orbiting and infalling terms, allowing a joint fit of interior properties, such as mass and concentration, and features of the infall region, such as the splashback radius. In addition to the density profile, we also include a mis-centering framework in the $\Delta\Sigma$ model \citep{Johnston07}. 

The largest difference between the profile fits is in the miscentering parameters. The effect of mis-centering is to dampen the signal in the inner region. Therefore, though the difference manifests in the mis-centering parameters, this could also indicate that we lack the framework to model the impact of dynamical selection on the lensing profile. Such an improved model will be the subject of a follow-up work.

We will compare our results to expectations from simulations in Section~\ref{sec:illustris}, to gain further insight on the dynamical state split. First, however, we should highlight the remarkable similarity of these observational results to the predictions of \cite{Xhakaj2022}, in which they measure the dependence of the lensing profile on secondary halo properties in simulations. In their Figure~3, they show the lensing profiles of halos in the MultiDark Planck 2 simulations \citep{Prada2012}, binned by properties such as centre-of-mass offset, concentration, accretion rate, and the scale factor at which the halo was half its present mass, $a_{1/2}$ (equivalent to $z_{50}$). We find the same strong trend in the observed profiles: the evolved subsample, which we expect to have a smaller $a_{1/2}$ (or larger $z_{50}$), has a larger lensing amplitude in the inner regions compared to the evolving sample. On the other hand, both subsamples have similar profiles in the outer region. Our results represent the first observational evidence for this behaviour on cluster scales in a large sample.

\subsubsection{Mass comparisons from lensing profiles and catalogues}
\label{sec:mass_comparison_wh24_y21}

We can also compare the mean masses for each sample derived from the weak lensing fits to means of the masses reported in \citetalias{wen2024} and \citetalias{yang2021} for the same sets of clusters. The abundance-matching method employed in \citetalias{yang2021} produces masses more consistent with our lensing masses, compared to the halo-stellar mass scaling relations used in \citetalias{wen2024}. The differences between the weak lensing and \citetalias{yang2021} mean mass are $-3.0\sigma$ and  $-0.6\sigma$ for the evolved and evolving samples respectively. Relative to the mean masses reported in \citetalias{wen2024}, the differences are $-10 \sigma$ and $-2.7\sigma$ (a more detailed discussion can be found in Appendix~\ref{app:WL_fits}).

 These findings justify our use of the masses reported in \citetalias{yang2021}. However, the cluster centres reported in \citetalias{yang2021}, which are defined through a luminosity-weighted average of the cluster members, produce lensing profiles with a high degree of miscentering. We find that the BCG catalogue from \citetalias{wen2024} produces more accurate centroids, evidenced by greater lensing signal in the central region. For this reason, we combine information from both catalogues, using the BCG from \citetalias{wen2024} as the cluster centre, and the mass reported in \citetalias{yang2021} as the total halo mass.
 
\section{Insight from Illustris-TNG simulations}
\label{sec:illustris}
Cosmological simulations provide us with the opportunity to compare optically identified galaxy clusters to dark matter halos identified in 3D. They also enable us to compare the mass accretion histories of cluster samples with selected using different criteria. We make use of the IllustrisTNG ``TNG300-1'' magnetohydrodynamical simulation \citep{marinacci2018illustris,naiman2018illustris,nelson2018illustris,pillepich2018illustris,springel2018illustris} to further explore the properties of evolved and evolving clusters.

\subsection{Sample selection}

Our cluster selection from the TNG300-1 simulation is based on the friends-of-friends halo catalogue provided by IllustrisTNG. For each halo, we identify the member galaxies and satellites from IllustrisTNG's subhalo catalogue. The subhalo catalogue is constructed using the \textsc{subfind} algorithm, which identifies gravitationally bound overdensities of particles within halos \citep{springel2001}. We define the stellar masses of these subhalos as the sum of the stellar particle masses within 30~kpc of the subhalo's most bound particle. We choose this aperture size of 30~kpc as the 3D stellar mass within this aperture from simulations is comparable to the measured mass within 2D Petrosian radius around galaxies that is frequently used in observational studies \citep{schaye2014eagle,pillepich2018illustris,Engler2021}. Every halo containing at least 10 member galaxies with stellar masses $M_*\geq10^{10}M_\odot$ within $R_{500c}$ of the cluster's most bound particle in 3D is included in the sample. While our optically identified cluster sample contains systems at a broad range of redshifts, as seen in Figure~\ref{fig:cl_properties_r_ur}, TNG300-1 data consists of snapshots at specific redshifts. We select a sample of simulated clusters at a redshift of $z=0.4$ to match approximately the mean redshift of the observed cluster sample.

\begin{figure*}
    \centering
    \includegraphics[width=1.8\columnwidth]{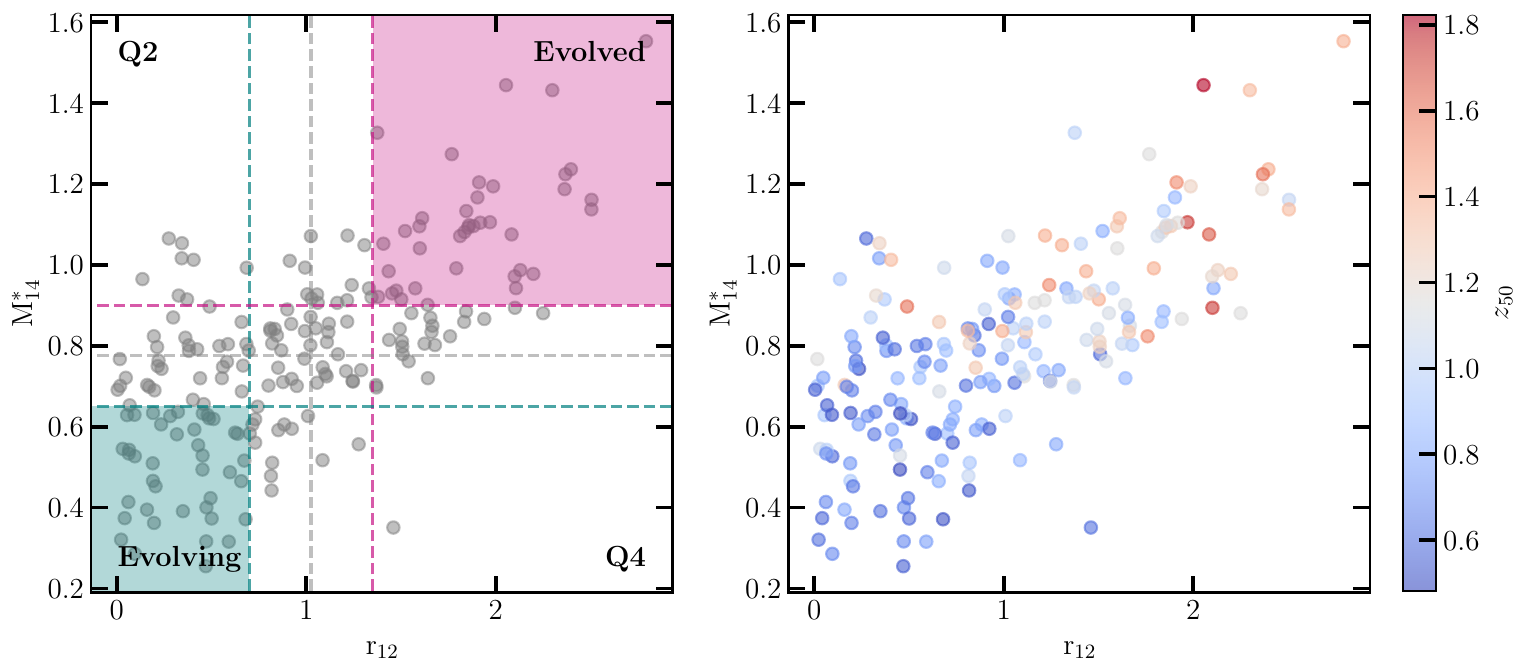}
    \caption{Cluster properties and assembly history at $z=0.4$ in the TNG300-1 simulation. The x-axes show the absolute $r$-band magnitude gap between the BCG and the brightest satellite galaxy of each cluster, and the y-axes show the log of the stellar mass ratio of the BCG to the 3rd most massive satellite of each cluster. The coloured dashed lines and shaded regions in the left panel indicate our selection regions for the ``evolved'' and ``evolving'' samples of simulated clusters. The grey dashed lines in the left panel indicate the regions defined as Q2 and Q4. The colour scale in the right panel indicates $z_{50}$, the redshift at which each cluster assembled half of its final mass.}
    \label{fig:TNG_magnitude_gaps}
 \end{figure*}

The stellar masses and r-band magnitudes of each TNG300-1 galaxy are computed from stellar particles within 30~kpc of the subhalo centres. The r-band magnitude gaps and the logarithm of the stellar mass ratio between the first and fourth most massive galaxies in each cluster are plotted for our full sample in Figure~\ref{fig:TNG_magnitude_gaps}. The selection of evolved and evolving halos in TNG300-1 is analogous to that performed for the DESI Legacy survey-based cluster sample. The $20\%$ of halos with the greatest magnitude gaps and stellar mass ratios are selected as ``evolved'' systems, while the $20\%$ of halos with the smallest magnitude gaps and stellar mass ratios are selected as ``evolving''. This division is indicated by the shaded regions in Figure~\ref{fig:TNG_magnitude_gaps}. The magnitude gaps of the TNG300-1 clusters span a slightly different range relative to the observed sample (shown in Figure~\ref{fig:cl_selection_criteria}). We adjust our selection boundaries in order to produce similar fractions of evolved and evolving halos in TNG300-1. The distributions of magnitude gaps in the observed and simulated samples are quite different, possibly because of their different redshift distributions. On the other hand, the distributions of stellar mass ratios between the two are quite similar.

In the right panel of Figure~\ref{fig:TNG_magnitude_gaps}, we present the redshift $z_{50}$ at which each halo had accumulated 50\% of the $M_{200c}$ mass it had at $z=0.4$. This is shown as a colour scale in the distribution of magnitude gaps and stellar mass ratios. The colour gradient across the distribution confirms that systems classified as evolving accreted the majority of their mass more recently, while the evolved systems accreted the majority of their mass at higher redshift. Clusters in the Q2 and Q4 quadrants have similar mean $z_{50}$ values, which indicates that the r-band magnitude gap and the first-to-fourth stellar mass ratio both carry similar information about the evolutionary state of clusters.

\begin{figure*}
    \centering
    \includegraphics[width=1.75\columnwidth]{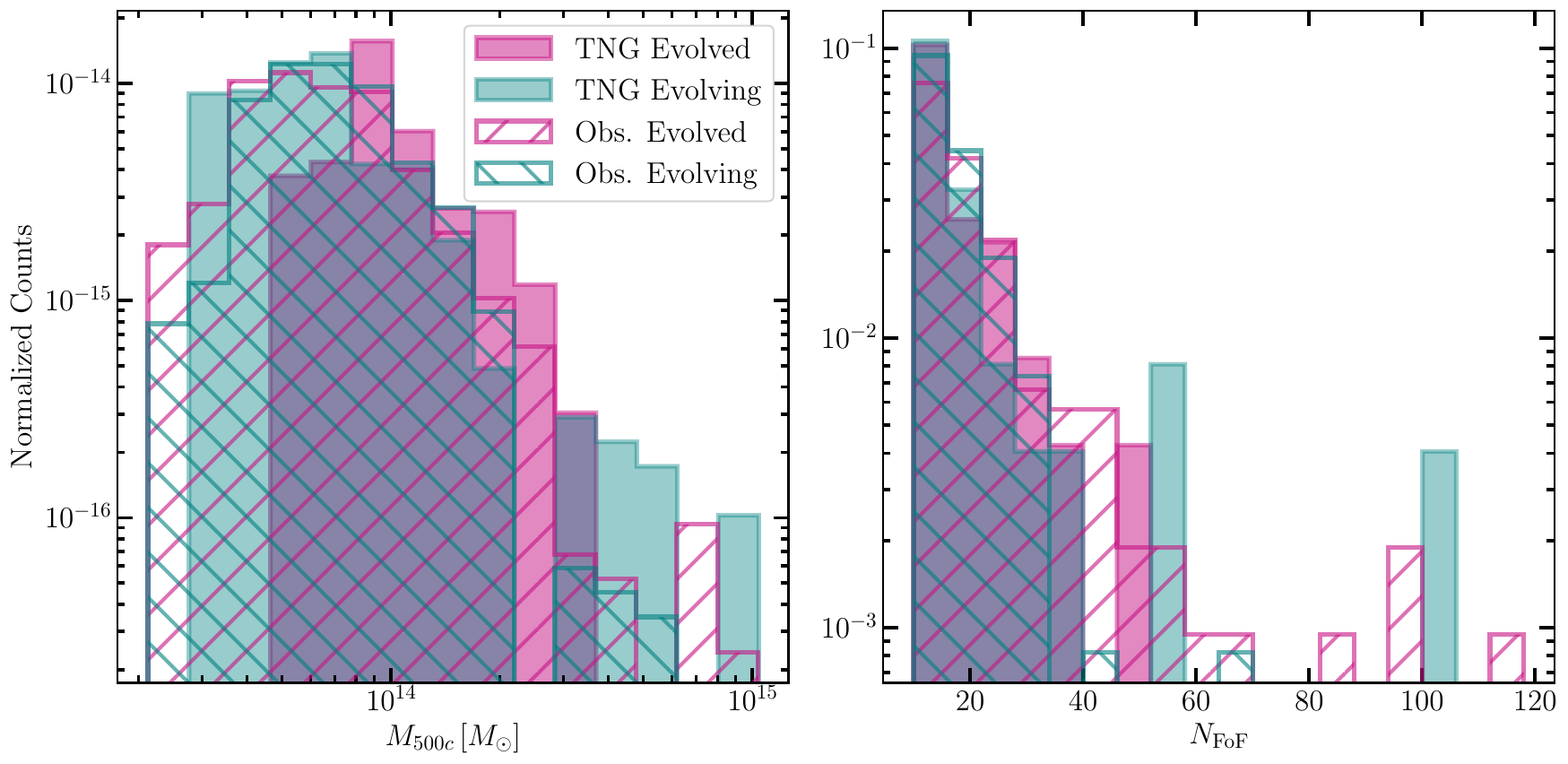}
    \caption{Comparison of the distributions of halo masses and number of member galaxies within $R_{500c}$ of the evolved and evolving samples from the TNG300-1 simulation and the observed samples.}
    \label{fig:TNG_cluster_masses}
\end{figure*}

The $M_{500c}$ mass distributions of the evolved and evolving TNG300-1 halos are presented in the left panel of Figure~\ref{fig:TNG_cluster_masses}. The cluster masses are those provided in the IllustrisTNG catalogue. We see that the evolved clusters have a greater average mass, while the evolving systems span a greater range of masses, including the most and least massive systems. The right panel of Figure~\ref{fig:TNG_cluster_masses} presents the distribution of the number of galaxies with stellar masses $M_\star\geq10^{10}\,M_\odot$ that lie within $R_{500c}$ in 3D. Similarly to the mass distribution, evolved clusters have a greater average richness, while evolving systems span a greater range of richnesses. These distributions contrast with the observed DESI Legacy sample, where evolved and evolving clusters have similar mass distributions but evolving clusters are slightly more massive and have greater richness on average (cf.~Figure~\ref{fig:cl_properties_r_ur}). Despite these differences between evolved and evolving clusters, the total halo mass distributions of both the optical survey clusters and the simulated clusters are similar. Far fewer clusters are identified in TNG300-1 compared to the observed sample, due to the simulation's limited volume. As a result, random scatter in the cluster mass and member galaxy distributions may account for some of the discrepancy between the evolved and evolving samples in TNG300-1.

\subsection{Stellar mass functions}
\label{sec:smfs_sims}

\begin{figure}
    \centering
    \includegraphics[width=\columnwidth]{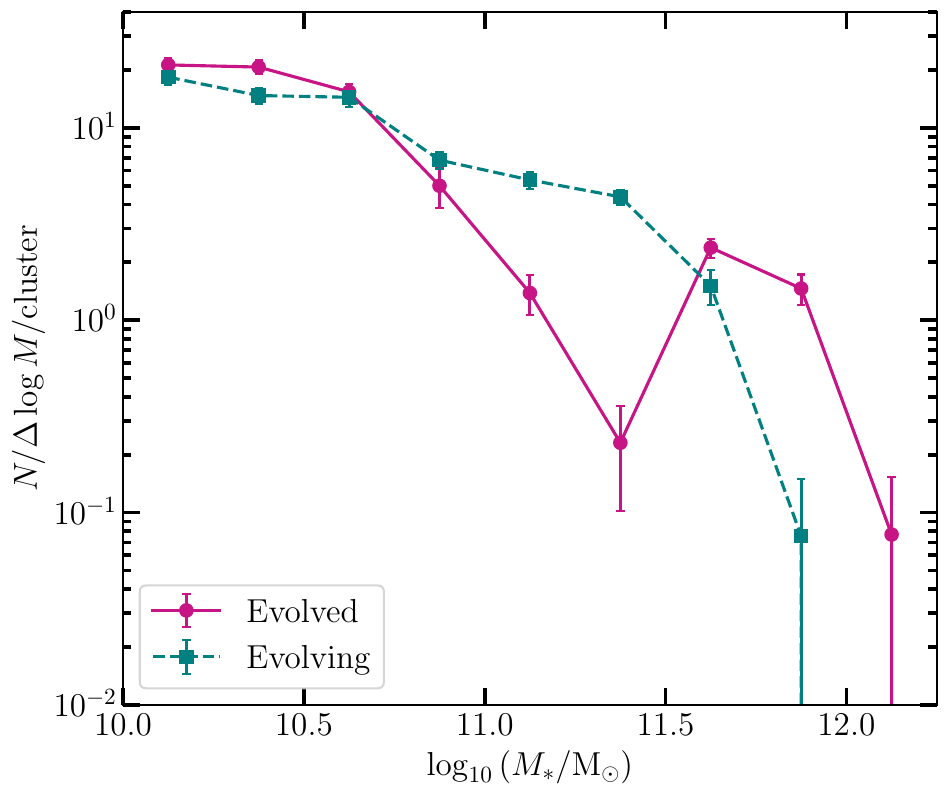}
    \caption{Stellar mass functions of the evolved and evolving samples from the TNG300-1 simulation at $z=0$.}
    \label{fig:TNG_stellar_mass_functions}
 \end{figure}

The mean stellar mass functions of our TNG300-1 samples of evolved and evolving clusters are presented in Figure~\ref{fig:TNG_stellar_mass_functions}. The stellar mass function for each cluster is computed from subhalos within $R_{500c}$ of the halo centre in 3D. The data points in Figure~\ref{fig:TNG_stellar_mass_functions} represent the mean number of galaxies per cluster in equally spaced $\log M_\star$ bins for our evolved and evolving cluster samples. The error bars represent the uncertainty on the mean for each sample. Consistent with the survey sample's stellar mass functions in Figure~\ref{fig:smf_r_ur_sat}, evolved clusters have a deficit of intermediate-mass galaxies and an excess of high-mass galaxies relative to evolving clusters. (In Figure~\ref{fig:TNG_stellar_mass_functions} we show mass functions in the $z=0$ snapshot, as this distinction is clearest at low redshift.) We expect these features to be consistent across observed and simulated systems, as in both cases evolved and evolving clusters are selected based on similar magnitude gaps and stellar mass ratios. This guarantees that evolved clusters have dominant BCGs and no similarly bright or massive satellites, while evolving clusters have several galaxies of mass and brightness similar to the BCG.

\subsection{Weak lensing signal}

While gravitational lensing signals are detected in weak lensing surveys via the shear of background galaxies, we numerically compute $\Delta\Sigma(R)$ directly from the mass distributions of halos in the TNG300-1 simulation. Given the full-volume particle data for the TNG300-1 simulation snapshot, we generate the lensing profiles around the most bound particle of each halo in our sample. The lensing profile of each halo is computed from a cylinder of particles with a radius of 20~cMpc tangent to the $z$-axis and a $\pm20$~cMpc depth along the $z$-axis. We found that the lensing profiles are not significantly altered by increasing the line-of-sight depth along the z-axis by a factor of 2. Given the profiles around individual halos, we compute the mean lensing profiles of our selected groups of clusters.

\begin{figure}
    \centering
    \includegraphics[width=\columnwidth]{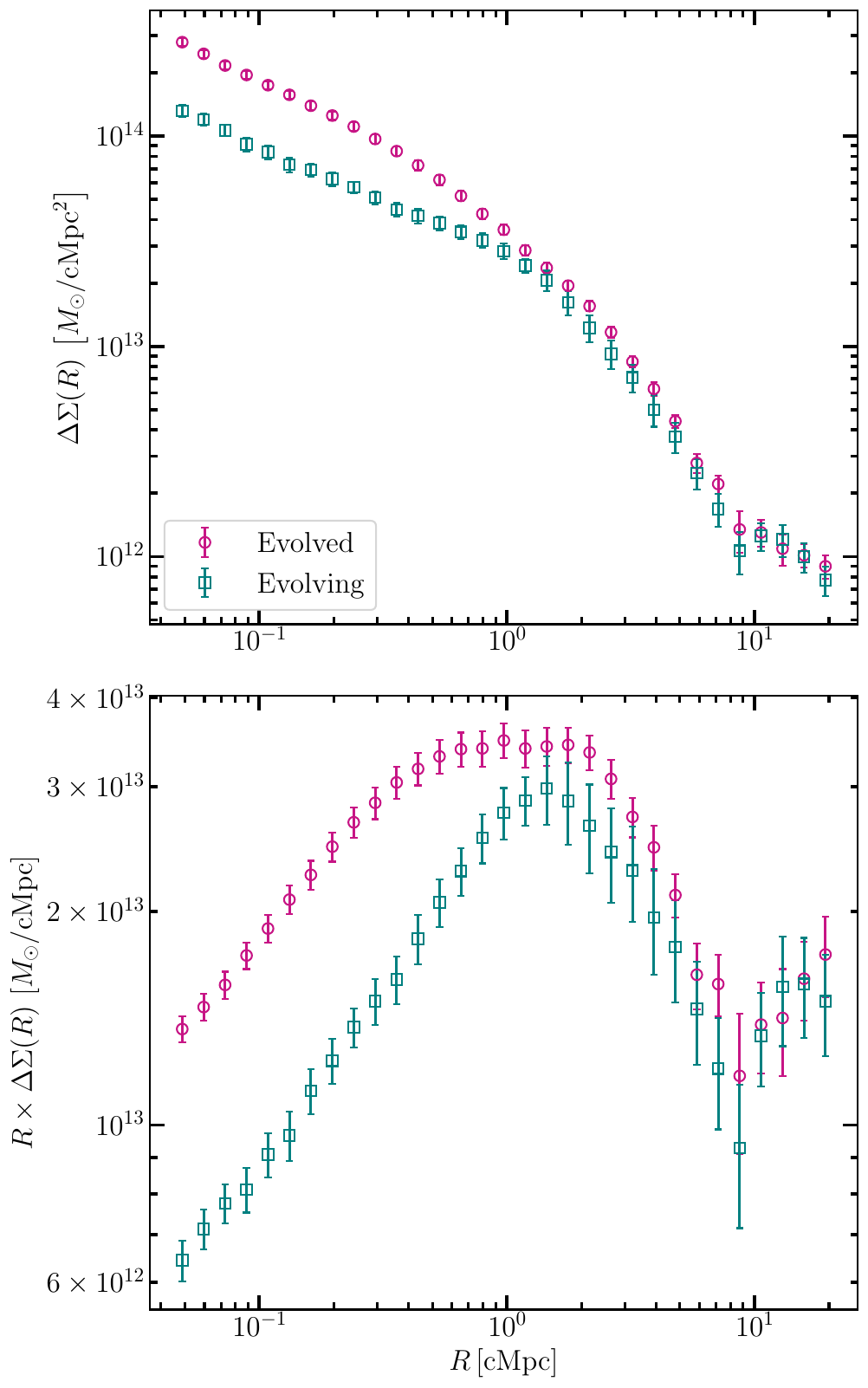}
    \caption{Mean weak lensing $\Delta\Sigma$ profiles of the evolved and evolving samples from the TNG300-1 simulation. The profiles are computed directly from the particle mass distribution of each halo rather than from the shear of background galaxies. The bottom panel scales $\Delta\Sigma$ by the projected radius from the cluster centre.}
    \label{fig:TNG_lensing}
 \end{figure}

In Figure~\ref{fig:TNG_lensing}, we show the mean lensing signals from the evolved and evolving TNG300-1 cluster samples. The error bars represent the uncertainty on the mean signal in each radial bin. At large radii, the two samples have similar amplitudes, indicating that evolved and evolving clusters exist in similarly dense large-scale environments. 
At intermediate radii, the amplitude of the profile is also slightly ($\sim\,$20\%) higher, consistent with the greater mean mass of the evolved sample discussed previously. On the other hand, at radii of less than 1 cMpc, there is a clear difference in the mean profiles of the two samples, with evolved clusters having significantly higher central densities. 

Unlike those of observed clusters, the lensing profiles of simulated TNG300-1 halos are not subject to miscentering, each $\Delta\Sigma$ profile being centred on the most bound particle in each cluster. Thus, the difference in the inner lensing profiles reflects a genuine structural difference in the mean mass distributions of the two samples.
This further validates the difference between the mean lensing profiles of the two observed samples shown previously in Figure~\ref{fig:lensing_profiles}.

\subsection{Relation to assembly history}

\begin{figure}
    \centering
    \includegraphics[width=0.99\columnwidth]{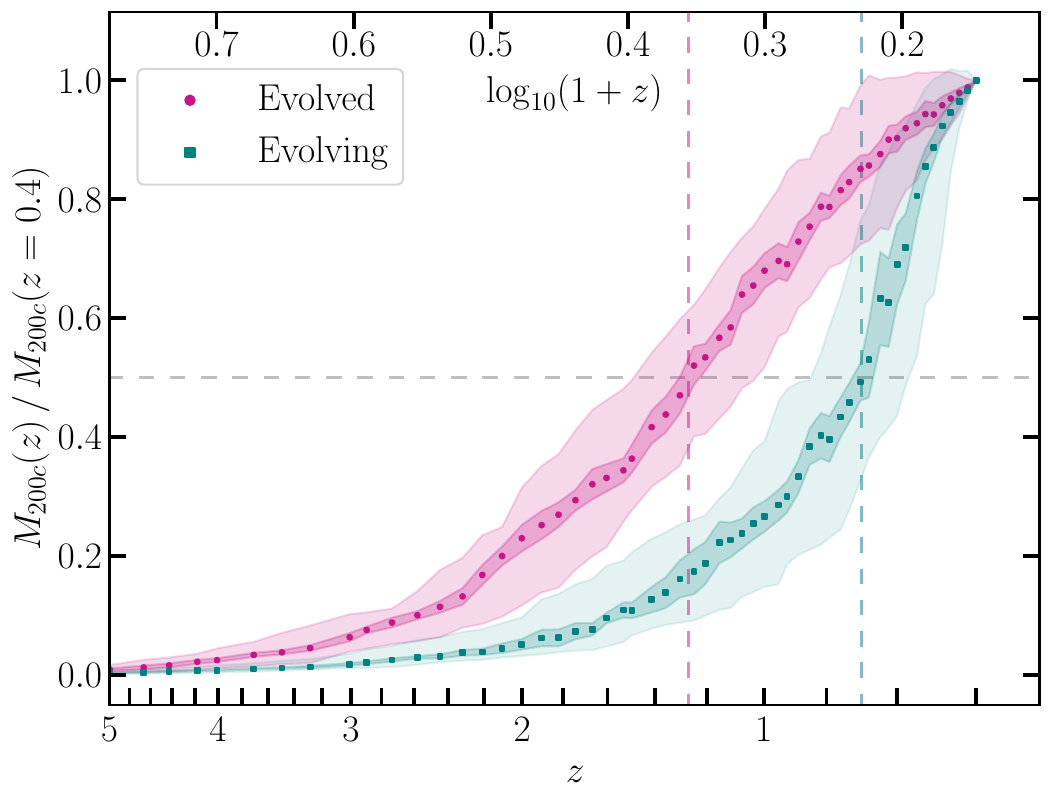}
    \includegraphics[width=\columnwidth]{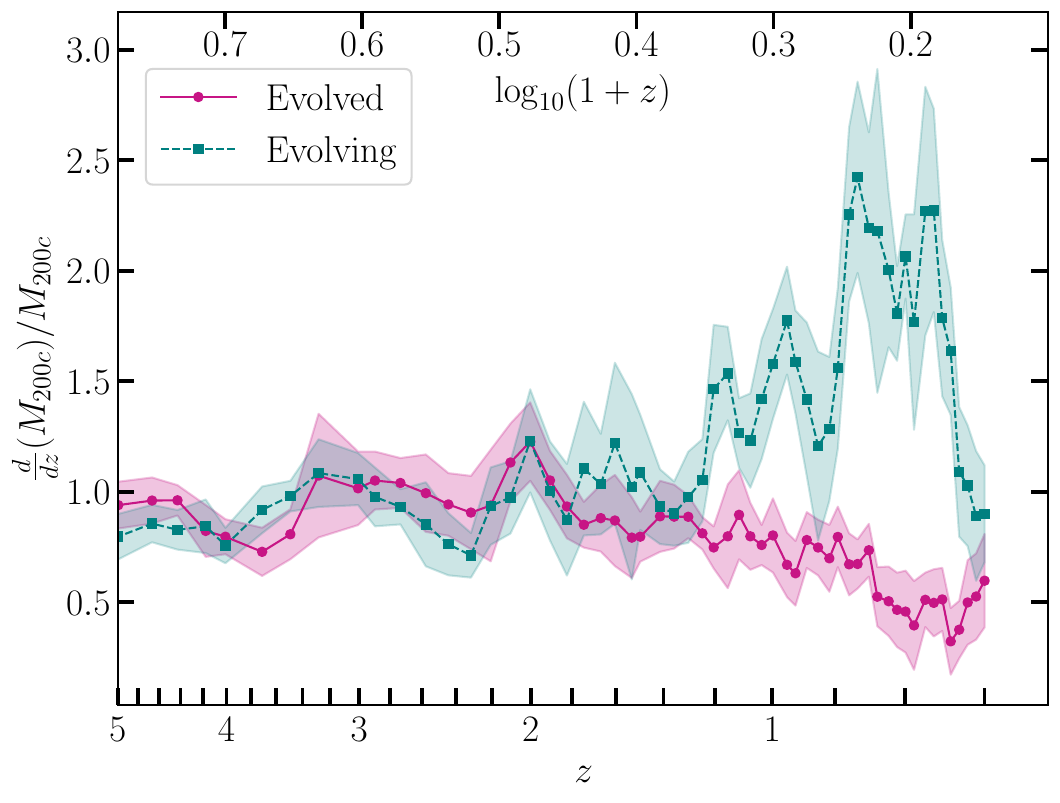}
    \caption{The median mass accretion histories of the evolved and evolving samples from the TNG300-1 simulation. In the top panel, the points represent the median masses of halos at each available simulation snapshot. The thin shaded regions represent the uncertainties on the medians. The large shaded regions represent the $1\sigma$ scatter around the medians. The dashed lines indicate the average formation redshift $z_{50}$ for each sample. In the bottom panel, the points represent the median mass accretion rates with respect to redshift. The shaded regions represent the errors on the medians.}
    \label{fig:TNG_growth_history}
\end{figure}

In the top panel of Figure~\ref{fig:TNG_growth_history}, we present the median normalised mass accretion histories of the evolved and evolving samples of TNG300-1 halos. The dashed lines show the mean redshift $<z_{50}>$ by which halos in each sample had assembled 50\% of the mass they reached at $z=0.4$. For evolved halos, $<z_{50}> \sim 1.3$, while for  evolving halos $<z_{50}> \sim 0.75$. The wide separation of the median mass accretion histories of evolved and evolving halos relative to the scatter within each sample (indicated by the shaded regions) emphasizes how effectively magnitude gaps and stellar mass ratios may be used to define cluster samples with different evolutionary states. In the bottom panel of Figure~\ref{fig:TNG_growth_history}, we present the median normalised mass accretion rates of the evolved and evolving samples. These are computed for each halo by taking a numerical derivative of $M_{200c}(z)$ with respect to redshift. Both samples experience steady accretion from $z=6$ to $z\simeq2$, at which point evolved systems experience a slight decrease in their accretion rate, while evolving systems experience a large increase. Overall, evolving halos have accreted much more mass at low redshift, which affects their structure at the observed redshift. 

Previous work by \citet{Xhakaj2022} compared the gravitational lensing profiles of simulated galaxy cluster samples selected based on their mass accretion rates within the last dynamical time. The mass accretion rate used by \citet{Xhakaj2022} is defined by \citet{Diemer2017} as:
\begin{equation}
    \Gamma_\mathrm{dyn}(t)=\frac{\log[M_{200m}(t)]-\log[M_{200m}(t-t_\mathrm{dyn})]}{\log[a(t)]-\log[a(t-t_\mathrm{dyn})]}\,,
    \label{eq:accretion-rate}
\end{equation}
where $t_\mathrm{dyn}$ is the dynamical time, $a$ is the scale factor, and $M_{200m}$ is the mass of the halo within a radius where the average mass density is 200 times the mean density of the universe. The dynamical time is defined by \citet{Diemer2017} as the crossing time for a particle through a halo with mean density 200 times the mean density of the universe, which reduces to:
\begin{equation}
    t_\mathrm{dyn}=\frac{1}{5\sqrt{\Omega_m}}\cdot\frac{1}{H(z)}\,,
    \label{eq:dynamical-time}
\end{equation}
where $\Omega_m$ is the matter fraction and $H(z)$ is the Hubble parameter. \citet{Xhakaj2022} found that in simulations, observational proxies for $\Gamma_\mathrm{dyn}$ with scatter $\sigma_{\Gamma_\mathrm{dyn}|\mathrm{obs}}<1.5$ could be used to select subsamples with different growth histories and detect significant differences in their gravitational lensing profiles, given the lensing data anticipated from the full-area HSC survey \citep{aihara2018,aihara2019}.

\begin{figure}
    \centering
    \includegraphics[width=\columnwidth]{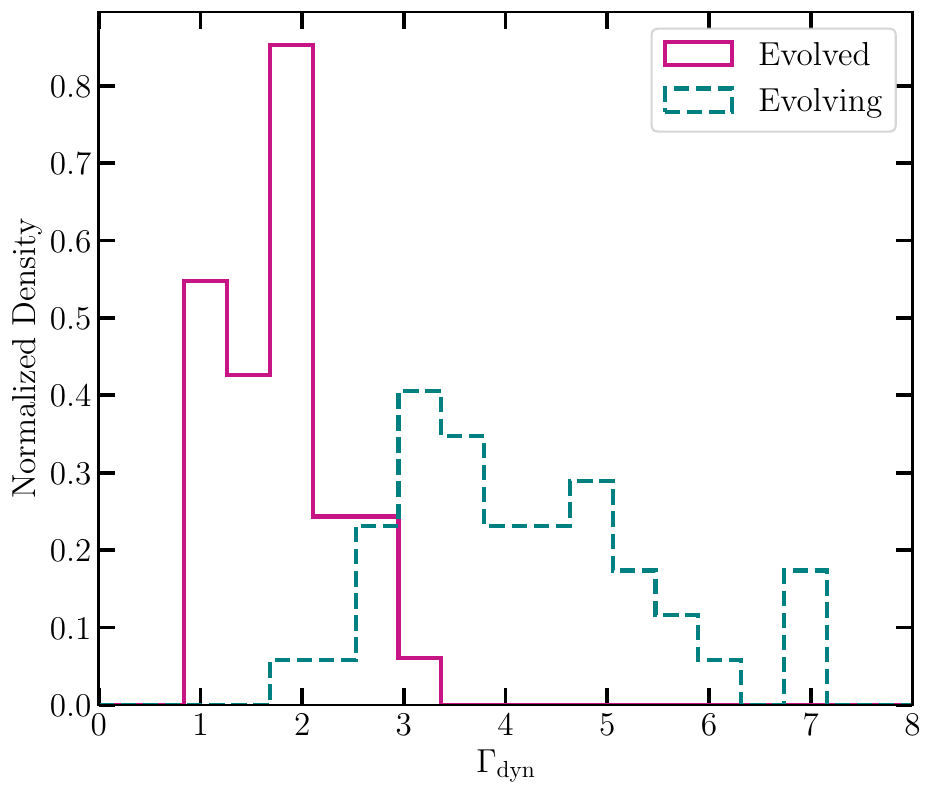}
    \caption{The mass accretion rates over the last dynamical time of evolved and evolving halos from the TNG300-1 simulation.}
    \label{fig:TNG_accretion_rate}
\end{figure}

In Figure~\ref{fig:TNG_accretion_rate}, we compare the mass accretion rates $\Gamma_\mathrm{dyn}$ of our evolved and evolving TNG300-1 cluster samples. There is little overlap in the distributions of mass accretion rates for the two samples: 87\% of evolved halos have $\Gamma_\mathrm{dyn}<2.5$, while 95\% of evolving halos have $\Gamma_\mathrm{dyn}>2.5$. The scatter $\sigma_{\Gamma_\mathrm{dyn}|\mathrm{obs}}$ is less than 1.5 for both samples, and thus our indicators of dynamical state are sufficiently accurate that we should expect significant differences in the lensing profile, consistent with the results shown in Figure~\ref{fig:TNG_lensing}. (See Figure 7 from \citet{Xhakaj2022} for comparison.)

\section{Discussion}
\label{sec:discussion}

Our tests using the Illustris TNG300-1 simulations in Sec.~\ref{sec:illustris} clearly demonstrate that the selection of evolved and evolving samples based on the magnitude gap and stellar mass ratio separates the samples in their formation history in terms of mass accretion. In this section, we explore how the mass accretion history can be translated into the magnitude gap and mass ratios.

\subsection{Stellar mass fraction and cumulative mass fractions}
\label{sec:smf_density_and_cumulative_smf}

\begin{figure*}
    \centering
    \includegraphics[width=1.8\columnwidth]{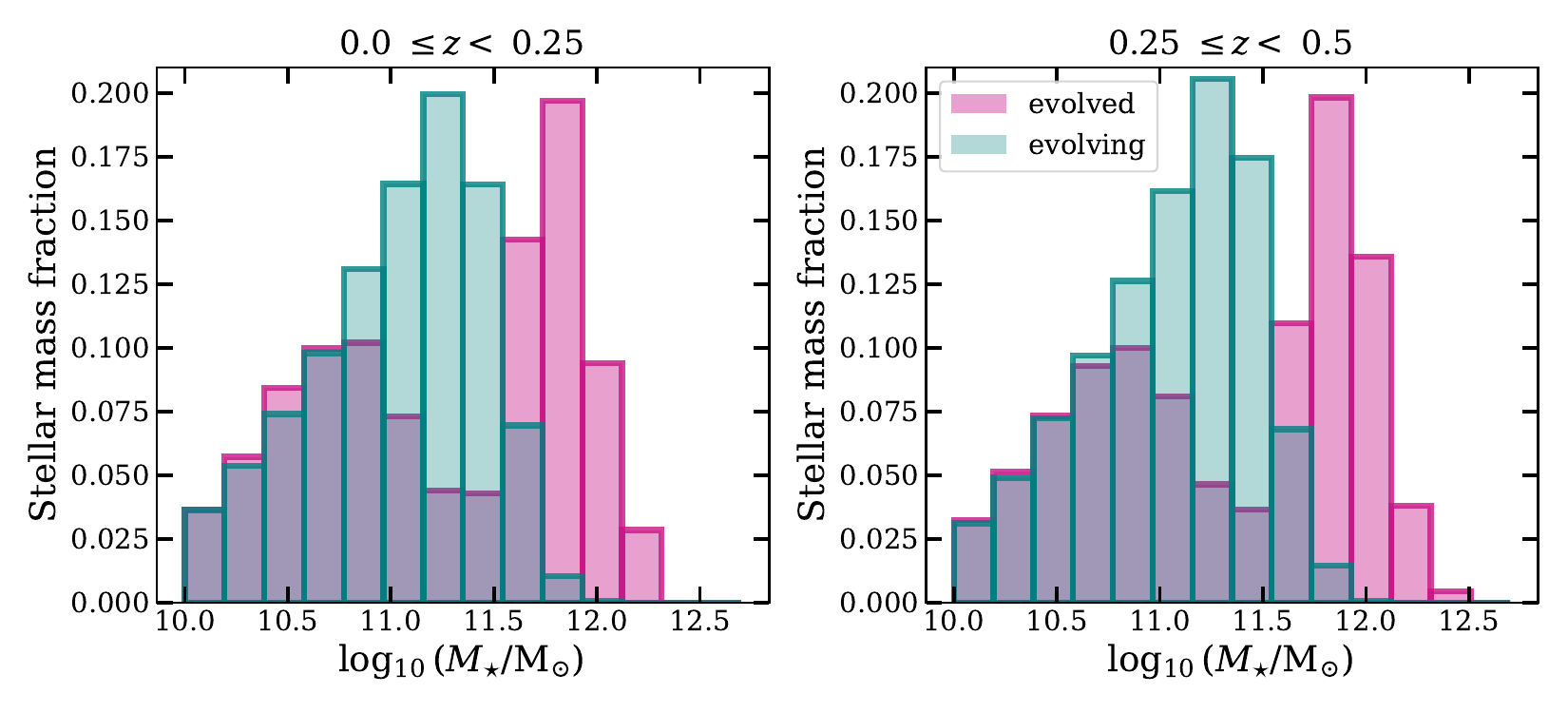}
    \caption{Stellar mass fraction of our evolved (pink) and evolving (teal) cluster samples, split by two redshift bins from left to right panels. The highest contribution to the total cluster mass is coming from galaxies with $\mstar \approx 10^{11.3}\msun$, and $\mstar \approx 10^{12.0}\msun$ in evolving and evolved clusters, respectively.}
    \label{fig:smf_den_r_ur}%
 \end{figure*}

The mean SMF showed a clear difference between evolved and evolving samples, but no strong redshift dependence for either sample (Sec.~\ref{sec:r_ur_smf}). Fig.~\ref{fig:smf_den_r_ur} shows how stellar mass is distributed between galaxies of different masses in each sample. For this, we take the stellar mass weighted SMF for evolved and evolving systems, and normalise it using the following equation:

\begin{equation}
    \mathrm{Stellar\ mass\ fraction}_{state} = \frac{M_\star \times N(M_\star)}{\Sigma M_{\star, {state}}}
\label{eqn:sm_fraction}
\end{equation}

where $N(M_\star)$ is the SMF, and $state$ is either evolved or evolving. In this formulation, the integral of each of the evolving and evolved mass fraction histograms in each redshift from Fig.~\ref{fig:smf_den_r_ur} equals unity. In evolving systems, galaxies with $\mstar \approx 10^{11.3}\msun$ contribute the greatest fraction of the stellar mass. There are almost no galaxies with $\mstar \approx 10^{12.0}\msun$, and they contribute negligibly to the total stellar mass. In evolved systems, on the other hand, galaxies with $\mstar \approx 10^{12}\msun$ contribute the greatest fraction of the stellar mass, while there is a clear drop in the stellar mass fraction at intermediate galaxy mass.

Accretion onto a cluster should eventually lead to the growth of the central galaxy, either through slow, smooth accretion of diffuse material, or through rapid mergers with satellites. Thus, in going from evolving to evolved systems, we expect a net increase in the mean stellar mass of the BCG at fixed halo mass. The lack of $\mstar \approx 10^{11.3}\msun$ galaxies in evolved systems compared to that in evolving systems, and the similarity of contributions to the total mass from lower mass galaxies ($\mstar \leq 10^{11.0}\msun$), suggests that the growth of the BCG comes primarily from high-mass satellites rather than lower mass ones. One caveat is our lack of data at the low-mass end of the mass function; as a result, we cannot say anything about whether dwarf galaxies or $\mstar \approx 10^{11.3}\msun$ galaxies contribute more to the build-up of the BCG. Previous work on BCG growth in clusters shows that the contribution from dwarf galaxies and Milky Way-like galaxies is non-negligible \citep[e.g. ][]{kluge2024,Demaio2020}. Another missing piece in understanding the growth of the BCG is the total amount of diffuse intracluster light (ICL). Recent studies have found consistent signatures of co-evolution between BCG and ICL, highlighting the need to consider BCG growth alongside that of the ICL, as discussed further below.

 \begin{figure*}
    \centering
    \includegraphics[width=1.8\columnwidth]{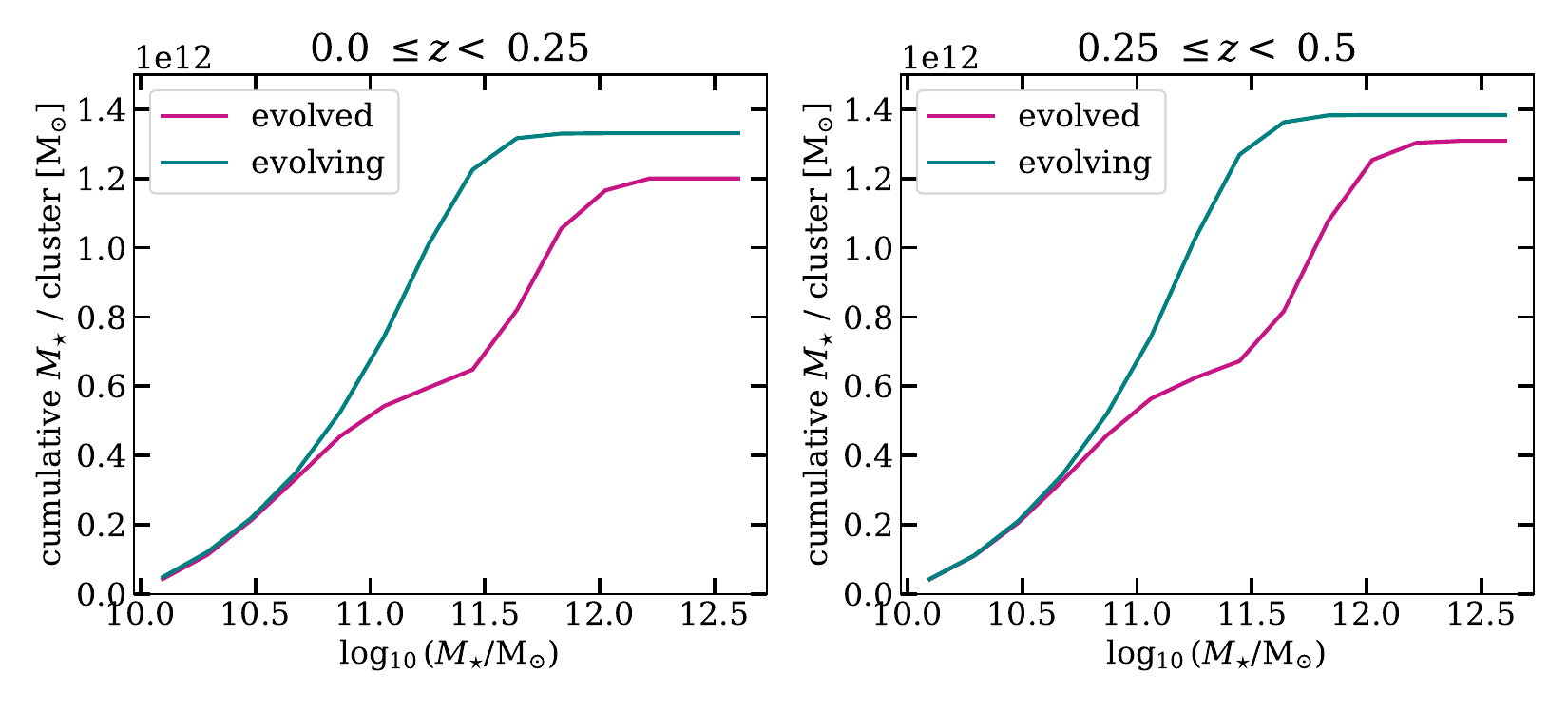}
    \caption{Cumulative stellar mass of our sample per cluster in two redshift bins from left to right panels, split by evolved (purple) and evolving (teal) states. The final cumulative mass for each subsample is the total stellar mass in an average cluster in that sample.}
    \label{fig:cumulative_smf_r_ur}%
 \end{figure*}

Fig.~\ref{fig:cumulative_smf_r_ur} shows the mass fractions accumulated with increasing galaxy mass. 
The cumulative contributions in both evolved and evolving clusters are almost the same up to $\mstar \leq 10^{11.0}\msun$, but diverge at intermediate masses. The average total stellar mass is slightly higher ($\sim 10$\%) for the evolving systems, which is approximately consistent with their halo mass distribution, as shown by Fig.~\ref{fig:cl_properties_r_ur}.  
A possible missing link here is the ICL. Because of its diffuse nature, the ICL is difficult to measure in individual systems and is often excluded from BCG stellar mass estimates. Recent studies have shown that the fraction of stellar mass in ICL depends on the dynamical state and formation state of a cluster \citep{jimenezTeja2018,contini2023,kimmig2025}, with actively forming clusters having less ICL and evolved clusters having more ICL at the same halo mass and redshift. In a future work (Ahad et al. in prep.), we will measure the ICL content in our samples and test whether the ICL can explain the net stellar mass difference between them.  

\subsection{Radial red fraction profiles}
\label{sec:red_fractions}

\begin{figure}
    \centering
    \includegraphics[width=\columnwidth]{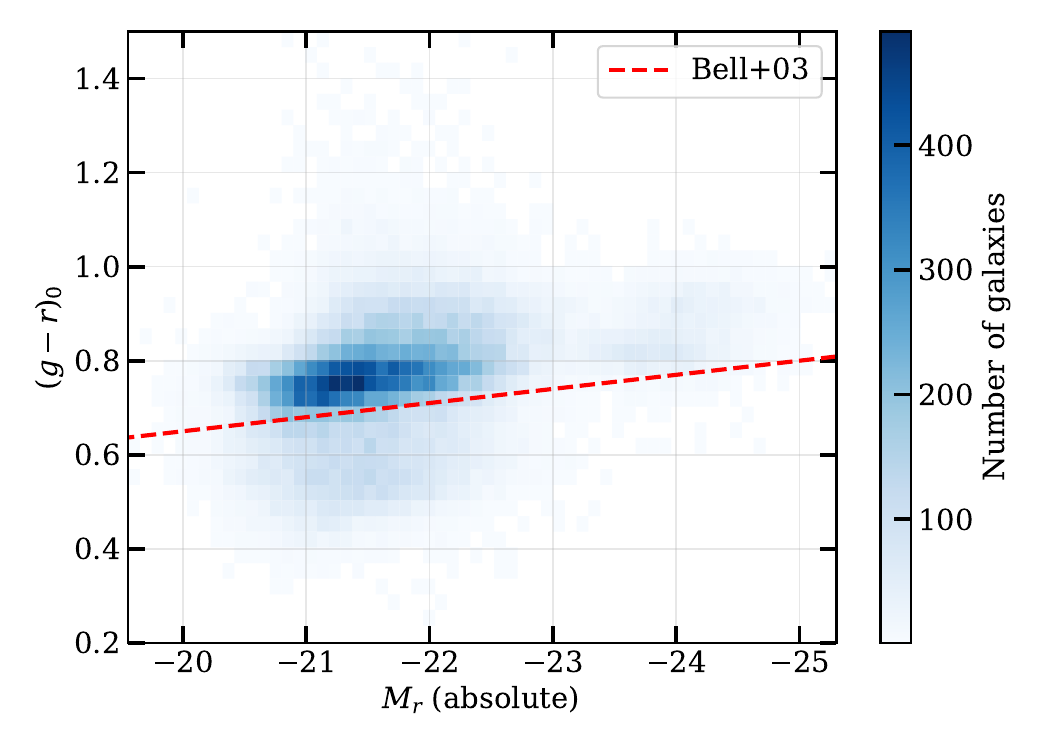}
    \caption{Rest frame $g-r$ colour vs.~absolute $r-$band magnitudes of all the cluster member galaxies within $R_{500}$. The red dashed line shows the threshold used to separate red and blue galaxies from \citet{Bell2003}. More details on the threshold selection are provided in the text.}
    \label{fig:CMD_red_selection}%
 \end{figure}

Given colour information about the member galaxies, we can also check for differences in the red fraction between our two samples. We use the k-corrected rest-frame $g-r$ colours and absolute $r-$band magnitude $M_r$ of the cluster member galaxies and split them between red and blue galaxies using the separation line given by $0.545 - 0.03\times(M_r + 16.5)$, as shown in Fig.~\ref{fig:CMD_red_selection}. This separation is based on the $g-r$ colours of galaxies from the Sloan Digital Sky Survey by \citet{Bell2003}. Figure~\ref{fig:CMD_red_selection} also shows that the majority of the cluster members are red. This is consistent with the literature that clusters host a higher fraction of red galaxies \citep{dressler1980ApJ...236..351D, balogh1999c, kauffmann2004environmental, weinmann2006properties, blanton2005relationship, Peng_2010, wetzel2012galaxy, woo2012dependence,hennign2017}. Photometric redshift-based cluster finding and the fact that we exclude lower mass galaxies ($\mstar \leq 10^{10.0}\msun$) also contribute to a high red fraction, as lower mass galaxies tend to have a lower red fraction. 

\begin{figure}
    \centering
    \includegraphics[width=0.9\columnwidth]{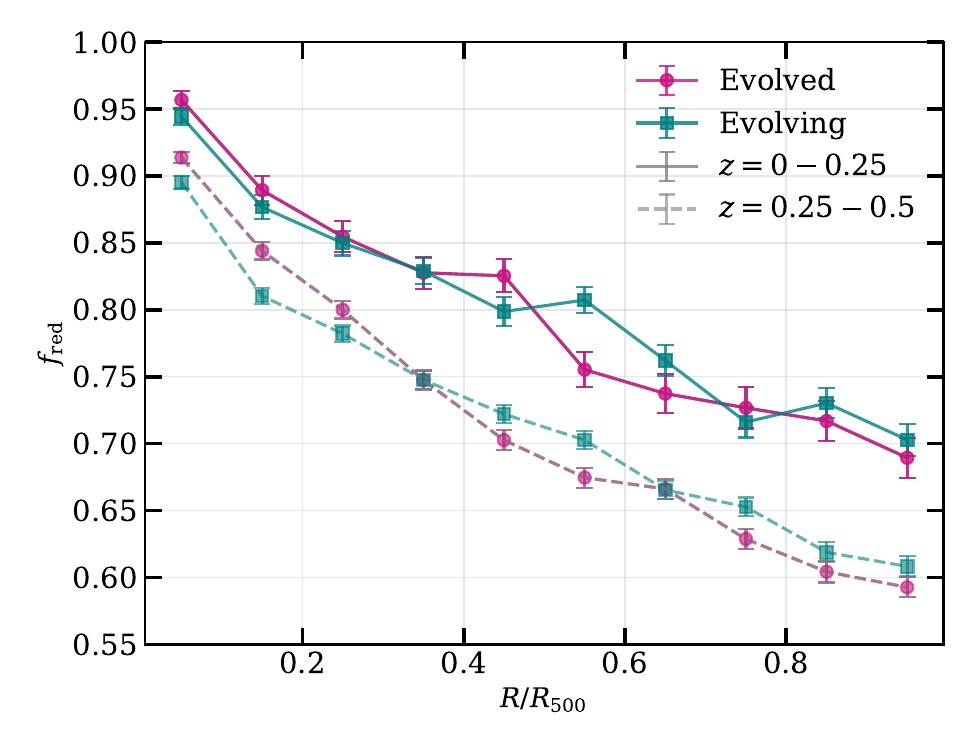}

    \includegraphics[width=0.9\columnwidth]{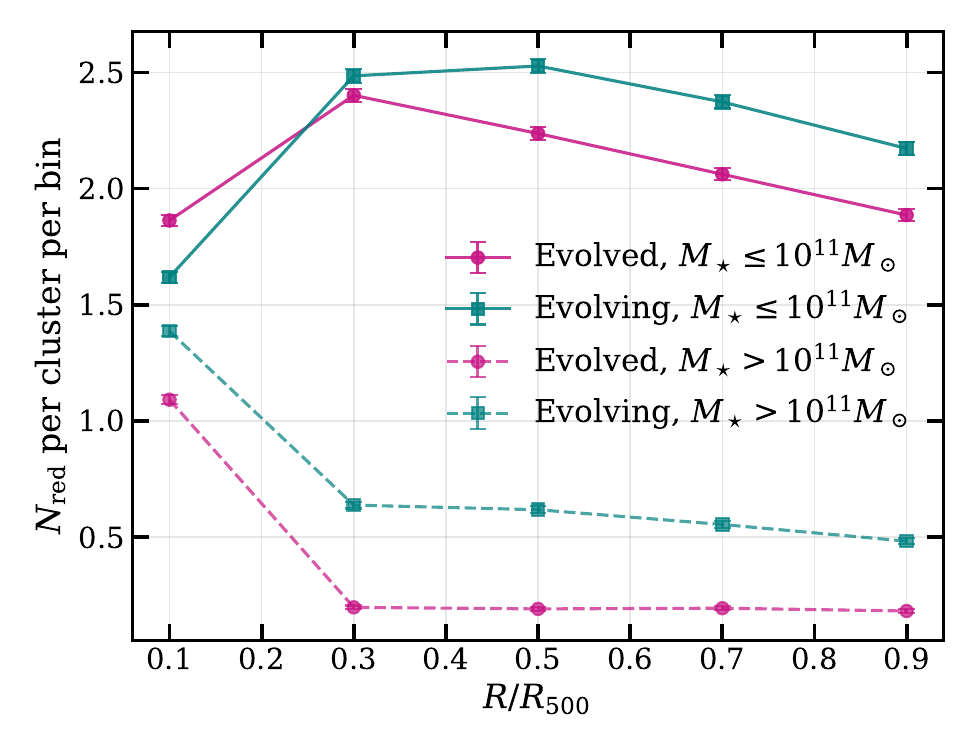}

    \caption{
    Top: Radial profiles of red fraction (quenched fraction) in the evolved and evolving clusters. The radial distance from BCG is scaled to the $r_{500}$ of each cluster before stacking them. Solid and dashed lines indicate two different redshift bins, and error bars show the $1$-$\sigma$ distribution of the measurement with 100 bootstrap resampling. The red fraction profiles are comparable for the evolved and evolving samples at each redshift, but there is about 5\% enhancement in the red fraction at the lower redshift bin.
    Bottom: Average number of red galaxies per cluster in the same radial bins as the top panel for the evolved and evolving clusters. Here, clusters are not binned by redshift, but the cluster galaxies are binned by their stellar masses. Galaxies with $\mstar\leq10^{11}~\msun$ \& $\mstar>10^{11}~\msun$ are shown by solid and dashed lines, respectively.
    }
    \label{fig:radial_red_fraction}%
\end{figure}
Fig.~\ref{fig:radial_red_fraction} shows the radial red fraction profiles out to $R_{500c}$ for the evolved and evolving samples, in two redshift bins (top panel). At each redshift, the red fraction profiles of evolved and evolving clusters are comparable, with a slightly higher red fraction in the inner regions of the evolved clusters. However, there is a clear trend with redshift, with at least a 5\% increase in the red fraction at all radii in the lower redshift bin. The similar red fraction in the evolved and evolving samples indicates that the difference in their SMF cannot be explained by the red galaxy fraction in these systems. 

The bottom panel of Fig.~\ref{fig:radial_red_fraction} shows the average count of red galaxies in the evolved and evolving clusters across all redshifts together. Cluster galaxies are divided into two bins according to their stellar mass ($\mstar\leq10^{11}~\msun$ \& $\mstar>10^{11}~\msun$). The average total number of galaxies is similar for the evolved (16.7) and evolving (20.1) samples, so the radial number counts demonstrate a relative central concentration (at $<0.2R_{500}$) of low-mass red galaxies in evolved clusters, as well as a deficit of massive red galaxies. The higher number of massive red galaxies in the central regions of evolving clusters is likely due to their selection based on the magnitude gap. There is also a higher count of red galaxies in the evolving clusters for both galaxy mass bins beyond $0.3R_{500}$, which is not connected to the selection process. A higher red galaxy count, combined with a similar red fraction in the outer region of the evolving clusters, indicates that the evolving clusters contain merging or recently merged groups/clusters that have a similar red fraction and a higher number of galaxies. These similar galaxy populations indicate that the massive galaxy populations in the evolving systems are already in place before they enter this state via preprocessing. This indication is consistent with recent work on galaxy properties in clusters of different dynamical states by \citet{veliz-astudillo2025b}.

\section{Conclusions}
\label{sec:conclusions}

The dynamical states and formation histories of galaxy clusters are expected to strongly influence the cluster galaxy populations. However, the impact of dynamical states is non-trivial to estimate because different indicators of cluster dynamical states can indicate different types of dynamical activity. Also, previous work in this area has been based either on simulations or on relatively small samples of observed clusters. Combining our large cluster sample and UNIONS lensing data, we can relate dynamical states more clearly to observable cluster properties. A summary of our approach and main findings is as follows:
\begin{itemize}
    \item We select subsamples of evolved and evolving galaxy clusters using the k-corrected absolute $r-$band magnitude gap between the BCG and the second brightest cluster galaxy, and the stellar mass ratio of the BCG and the fourth most massive cluster galaxy, since these are easily obtained from observations and known to relate to cluster growth history.
    \item The stellar mass functions (SMF) of the resulting evolved and evolving samples are distinct in several ways (Fig.~\ref{fig:smf_r_ur_sat}). Firstly, the SMF including BCGs has a double peak for the evolved sample, and a unimodal distribution for the evolving sample. Secondly, BCGs in the evolved sample are up to 0.5 dex more massive than those in the evolving sample. Finally, there is a lack of galaxies with $\mstar \approx 10^{11.3}\ \msun$ in the evolved clusters compared to the evolving clusters. The redshift evolution of the SMF is negligible. 
    \item The distribution of stellar mass vs.~cluster halo mass of the BCG, second brightest galaxy, and fourth massive galaxies show that at the same halo mass, BCGs in evolved systems are more massive, but second and 4th ranked galaxies are less massive compared to those in the evolving systems (Fig.~\ref{fig:BCG_M2_M4_mass_vs_halo_mass}). This observation suggests that the BCGs in the evolved systems grow from mass transfer from the most massive satellites. 
    \item The weak lensing mass profiles of these samples show that the evolving systems have a significantly reduced weak lensing signal at $<1$~Mpc around the cluster BCG (Fig.~\ref{fig:lensing_profiles}). By fitting a 3D density profile model to the lensing profiles, we find that the evolving sample contains clusters that are less concentrated (Fig.~\ref{fig:M_c}), have a larger degree of apparent mis-centering (Fig.~\ref{fig:PMC}), and have smaller, more pronounced splashback radii (Fig.~\ref{fig:rsp}).
    \item Using a similar selection method to create cluster samples in the Illustris TNG300-1 simulations, we find similar patterns for the SMFs (Fig.~\ref{fig:TNG_stellar_mass_functions}) and weak-lensing profiles (Fig.~\ref{fig:TNG_lensing}) between the simulated and observed clusters for both the evolved and evolving samples. The mean mass accretion histories of the evolved and evolving  samples differ significantly; the evolving sample experiences significant accretion over the past several Gyrs (Fig.~\ref{fig:TNG_growth_history}).
    \item The stellar mass fraction (Fig.~\ref{fig:smf_den_r_ur}) and cumulative contribution from galaxies of different stellar mass to the total cluster masses (Fig.~\ref{fig:cumulative_smf_r_ur}) support the idea that the BCGs in the evolved systems grow via mass transfers from massive satellites.
\end{itemize}

There are a few potential sources of uncertainty in this work. Firstly, the cluster catalogues we use are both based on photometric redshifts. The uncertainties in the photometric redshifts translate to a fairly broad redshift slice to identify cluster members. While galaxy colour selection still helps to improve the purity, there is always a chance of projected interloper galaxies being mistakenly identified as cluster members. In our current work, we used only high-mass galaxies ($\mstar \geq 10^{10}\ \msun$), where the photometric redshift uncertainties are smaller, but the chance of being a projected galaxy is still not zero. The presence of projection in photometrically selected galaxy clusters and the impact of projections on the cluster and galaxy properties is an active field of study \citep[e.g. ][]{erickson2011,busch2017,wojtak2018,sunayama2020,myles2021,lee2025}. A detailed study on the impact of projections in our photometric cluster catalogues will be presented in future work (Reid et al.~in prep.). 

Another limitation in our sample is the lack of low-mass galaxies, which may affect the red galaxy fraction measurements in this work. The photometric redshift selection can also be preferential to redder galaxies than bluer ones because the redshift uncertainties for bluer galaxies are larger \citep{navarro-girones2024}, and \citetalias{wen2024} removed galaxies with large redshift uncertainties. Thus, while the cluster member sample should be fairly complete at the stellar masses we include (as shown in \citetalias{wen2024}), it may still miss some bluer members due to the selection process. These problems can be reduced significantly by using spectroscopically selected cluster members, which will be possible with future data releases from the DESI survey \citep{desidr2025}. 

The above-mentioned uncertainties may affect the robustness of our conclusions, but they are unlikely to change any of the core findings. Overall, we show that subsamples with significantly different stellar and total mass distributions can be defined using a few simple, observable metrics. Given the systematic differences between the subsamples, general studies of cluster luminosity functions, mass functions, or lensing mass profiles should consider splitting their samples in a similar manner. The measured differences should also inform models of dynamical evolution. Finally, the dependence of cluster galaxy populations on cluster assembly history can be used to improve halo occupation distribution (HOD) models for future surveys. These findings also offer an observational link to assembly bias, demonstrating that the formation history of halos contributes to variations in galaxy populations and internal structure beyond what is captured by halo mass alone.

\begin{acknowledgments}
We thank members of the UNIONS collaboration for useful discussions. CTM is supported by an appointment to the NASA Postdoctoral Program at the NASA Goddard Space Flight Center, administered by Oak Ridge Associated Universities under contract with NASA. JET acknowledges support from the Natural Sciences and Engineering Research Council of Canada (NSERC), through a Discovery Grant. This research was enabled in part by support provided by Compute Ontario (www.computeontario.ca) and the Digital Research Alliance of Canada (alliancecan.ca). HH is supported by a DFG Heisenberg grant (Hi 1495/5-1), the DFG Collaborative Research Center SFB1491, an ERC Consolidator Grant (No. 770935), and the DLR project 50QE2305. MJH acknowledges support from NSERC through a Discovery Grant.

We are honored and grateful for the opportunity of observing the Universe from Maunakea and Haleakala, which both have cultural, historical and natural significance in Hawaii. This work is based on data obtained as part of the Canada-France Imaging Survey, a CFHT large program of the National Research Council of Canada and the French Centre National de la Recherche Scientifique. Based on observations obtained with MegaPrime/MegaCam, a joint project of CFHT and CEA Saclay, at the Canada-France-Hawaii Telescope (CFHT) which is operated by the National Research Council (NRC) of Canada, the Institut National des Science de l’Univers (INSU) of the Centre National de la Recherche Scientifique (CNRS) of France, and the University of Hawaii. This research used the facilities of the Canadian Astronomy Data Centre operated by the National Research Council of Canada with the support of the Canadian Space Agency. This research is based in part on data collected at Subaru Telescope, which is operated by the National Astronomical Observatory of Japan.
Pan-STARRS is a project of the Institute for Astronomy of the University of Hawaii, and is supported by the NASA SSO Near Earth Observation Program under grants 80NSSC18K0971, NNX14AM74G, NNX12AR65G, NNX13AQ47G, NNX08AR22G, 80NSSC21K1572 and by the State of Hawaii. 
\end{acknowledgments}

\begin{contribution}

The authors contributed to this work as follows. SLA led the project, prepared and analyzed the observational cluster sample, wrote and edited the manuscript, and produced the figures. RR conducted the analysis using the IllustrisTNG simulations, generated the corresponding figures, and wrote Section 5. CTM analyzed the weak-lensing profiles, prepared the relevant figures, and wrote Sections 4.3 and Appendix C. JET provided guidance on the overall direction of the project and contributed to writing the introduction and editing the manuscript. All authors reviewed and provided feedback on the final version of the paper.

\end{contribution}

\facilities{CANFAR, CADC}

\software{
Python (\href{http://www.python.org}{http://www.python.org}), \textsc{NumPy} \citep{Harris_et_al_2020}, \textsc{AstroPy} \citep{astropy2013}, \textsc{SciPy} \citep{jones2009}, \textsc{Matplotlib} \citep{hunter2007matplotlib}
}


\appendix

\section{Cluster dynamical state with X-ray information from eROSITA}
\label{app:erosita_comparison}

A widely used and independent indicator of cluster dynamical state is the offset between the centre of mass from cluster galaxies (often coinciding with the BCG) and the centre of hot intracluster gas from X-ray, also known as the centre of mass (CoM) offset. We use the eROSITA cluster catalogue from \citet{Bulbul2024} to compare the CoM offset with our evolved and evolving cluster selection.

\begin{figure}
    \centering
    \includegraphics[width=0.5\columnwidth]{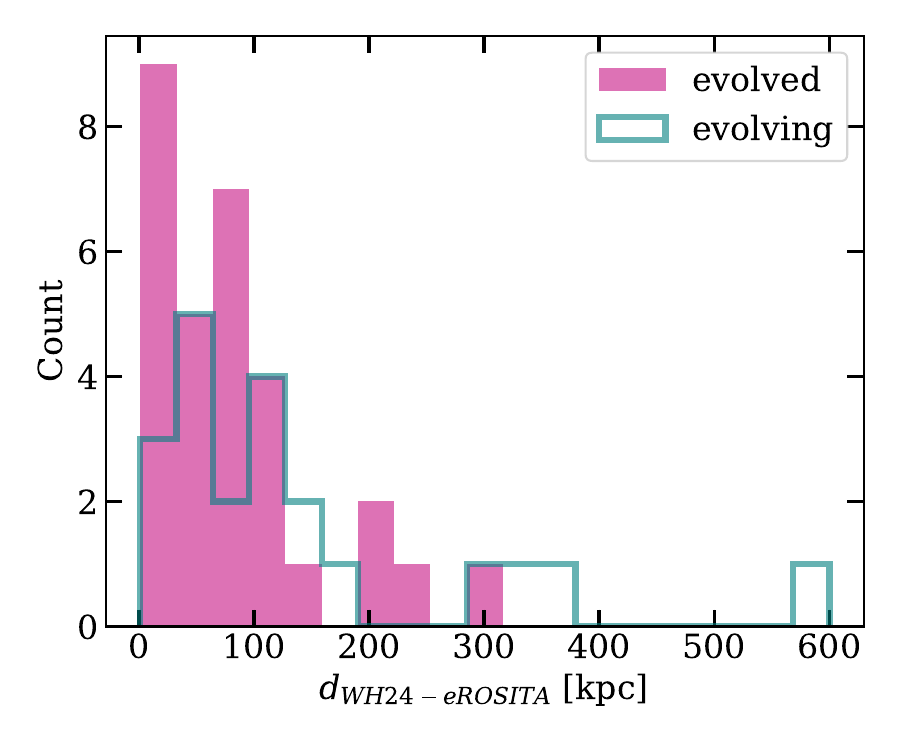}
    \caption{The radial distance in kpc between the centres of the cross-matched evolved and evolving clusters from eROSITA \citep{Bulbul2024} and \citetalias{wen2024} cluster catalogues. The offset between these two centres is less than 200~kpc for the majority of both types of clusters. However, all clusters with offsets of 300~kpc or more are identified as evolving clusters in our selection method, and the offset between cluster centres is smallest in the evolved sample.}
    \label{fig:distcen_erosita_wh24}
\end{figure}

The eROSITA catalogue covers mostly the southern sky; therefore, the overlap with our cluster sample from UNIONS, which primarily covers the northern sky, is small. Cross-matching our 3967 evolved and 3850 evolving clusters with the eROSITA cluster catalogue resulted in 30 and 21 matches, respectively. The distribution of radial CoM between the X-ray centres and BCGs from \citetalias{wen2024} is given in Fig.~\ref{fig:distcen_erosita_wh24}. The distributions show that most of the evolved clusters have CoM less than 100~kpc at their corresponding redshifts, and the peak of this distribution is at less than 20~kpc. Compared to that, the distribution of CoM of the evolving clusters are more spread out, with several having more than 300~kpc offset. Although the matched sample size is small, these distributions qualitatively agree with our evolved and evolving cluster selection from magnitude gap and stellar mass ratios. 

\section{Impact of large-scale environment on cluster properties}
\label{app:lss_impact}

To estimate the impact of large-scale environment of galaxy clusters in our sample, we take the galaxies within $1-10\times R_{500c}$ around our evolved and evolving cluster identified in Sec.~\ref{sec:lss_galaxy_search}, and calculate the surface stellar mass density within this area using the equation: $\sum_i(M_{\star,i})/(\pi*(R_2^2 - R_1^2))$, where $R_2 = 10R_{500c}$, $R_1 = R_{500c}$, and $M_{\star, i}$ is the stellar mass of all the galaxies found in this radial limit. We exclude the galaxies within $R_{500c}$ to remove correlation to the cluster mass. The average of the median surface mass density of the evolved and evolving clusters was used to split the clusters into dense and sparse large-scale environments. 

\begin{figure}
    \centering
    \includegraphics[width=0.6\columnwidth]{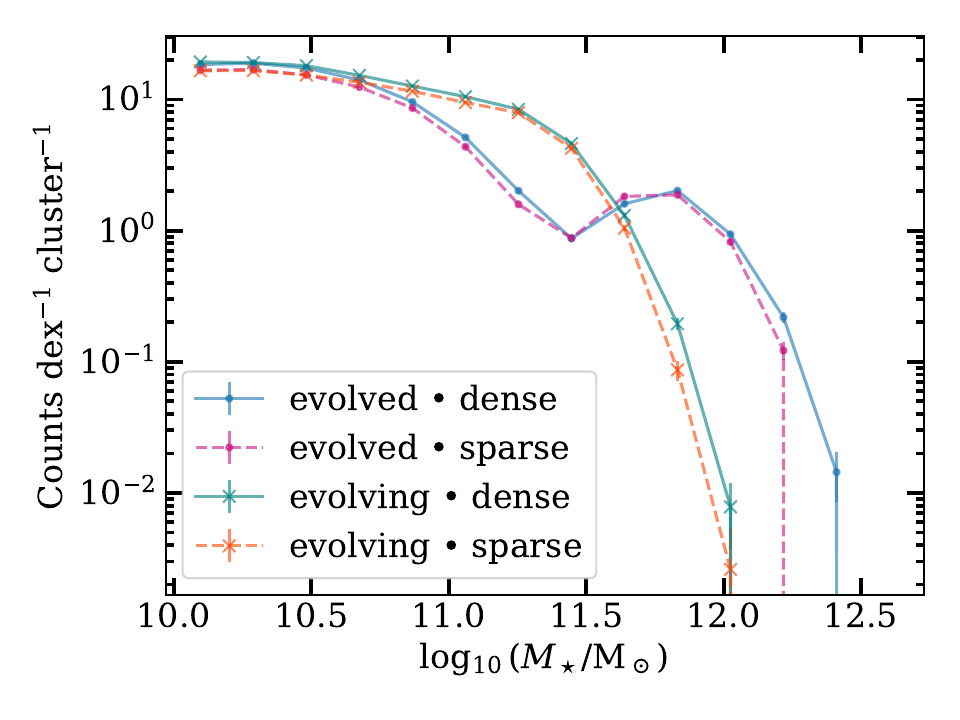}
    \caption{SMF of evolved and evolving clusters in dense and sparse large-scale environments. The evolved-evolving difference is bigger, but there is also a small signal in the dense-sparse separation.}
    \label{fig:smf_r_ur_d_s}%
 \end{figure}

\begin{figure*}
    \centering
    \includegraphics[width=0.9\columnwidth]{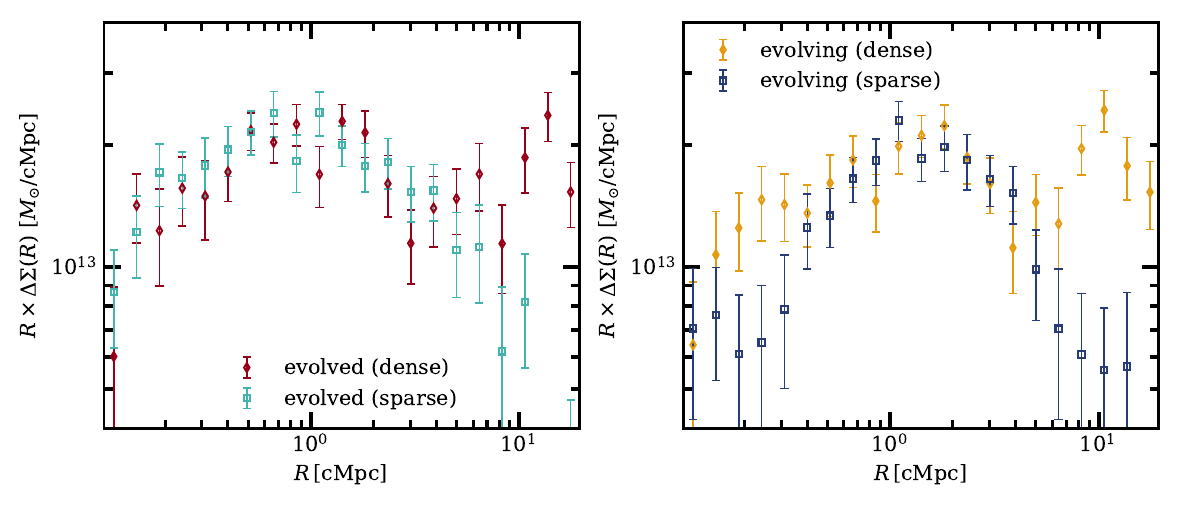}
    \caption{Lensing profiles of evolved and evolving clusters in dense and sparse large-scale environments. For both samples, there is a significantly enhanced (reduced) signal in dense (sparse) environments.}
    \label{fig:lens_ls_split}%
 \end{figure*}

The weak lensing profiles of the dense-sparse subsamples of the evolved and evolving clusters are given in Fig.~\ref{fig:lens_ls_split}. We find that the outer regions of these profiles ($>5\,$Mpc) show a significant difference between the dense and sparse subsamples. This difference means that in any sample of observed clusters, the outskirts of the lensing profiles may be able to indicate whether the ratio of dense/sparse split in the sample is even or skewed towards dense or sparse. One potential use of this insight is to compare multiple cluster catalogues, such as in \citet{Mpetha2025}. For the evolved sample, the inner region ($<1\,$Mpc) is not affected by the dense/sparse split, while for the evolving sample there is a very slight correlation. Overall, our dynamical state split is relatively independent of large-scale environment, and thus should be unaffected even if cluster selection is biased to higher or lower density regions.

The SMFs of samples split by environment do not show any strong difference either (Fig.~\ref{fig:smf_r_ur_d_s}). We note that the SMF includes only galaxies within $R_{500c}$ of the clusters, and thus this result is consistent with the lack of dependence on environment of the inner lensing profile, discussed above. Because galaxies in the outer regions are more likely to be affected by projection effects, we leave the exploration of the SMF at larger radii to future work.

\section{Weak lensing profile fitting for mass and splashback radius estimates}
\label{app:WL_fits}

To extract profile information for the evolved and evolving samples, we fit the 3D  density profile model introduced in \citet{Diemer2022} to the lensing profiles. It is given by
\begin{equation} 
        \rho(r) = \rho_{\rm orbit}(r) + \rho_{\rm infall}(r) \, ,
\end{equation}
where the orbiting term is a truncated Einasto profile,
\begin{equation}
    \rho_{\rm orbit}(r) = \rho_s \, \exp{\left(-\frac{2}{\alpha}\left[ \left(\frac{r}{r_s}\right)^{\alpha}-1\right]-\frac{1}{\beta}\left[\left(\frac{r}{r_t}\right)^{\beta}- \left(\frac{r_s}{r_t}\right)^{\beta}\right]\right)}  \nonumber \, ,
\end{equation}
and the infalling term is 
\begin{equation}
         \rho_{\rm infall}(r) = \rho_{\rm m}\,\left(1+ \delta_1/\sqrt{\left(\delta_1/\delta_{\rm max}\right)^2+\left(r/r_{\rm pivot}\right)^{2s}}\right) \, .
         \label{eq:rho_model}
\end{equation} 
We use an estimated $r_{200{\rm m}}$ for the pivot radius, converted from the mean catalogue $r_{500{\rm c}}$ using \textsc{colossus} \citep{Diemer2018} and the concentration model of \cite{Ishiyama2021}.

This density profile model is converted to $\Delta\Sigma$ when fitting the lensing profile, and a miscentering term is applied. There are two miscentering free parameters; the fraction of sources that have a centre offset from the true cluster centre, $f_{\rm off}$, and the average amplitude of this offset in units of cMpc, $\sigma_{\rm off}$. Thus, there are 10 free parameters in the $\Delta\Sigma$ model.

The fitting procedure is described in detail in Section~4.3 of \cite{Mpetha2025}. It involves performing multiple least squares regression fits, with initial conditions sampled over the prior range, then using the final best fit as input to a Monte Carlo sampler \citep{Karamanis2022a,Karamanis2022b}. As explored in \cite{Diemer2025,Mpetha2025}, fixing the two typically poorly constrained parameters $\alpha$ and $\delta_{\rm max}$ to sensible values reduces the variance on other parameters, while leaving the posterior means largely unchanged. We fix $\alpha$ and $\delta_{\rm max}$ to their best fit from an initial run, and then rerun the full fitting procedure now with 8 free parameters. 

\begin{figure}
    \centering
    \includegraphics[width=\columnwidth]{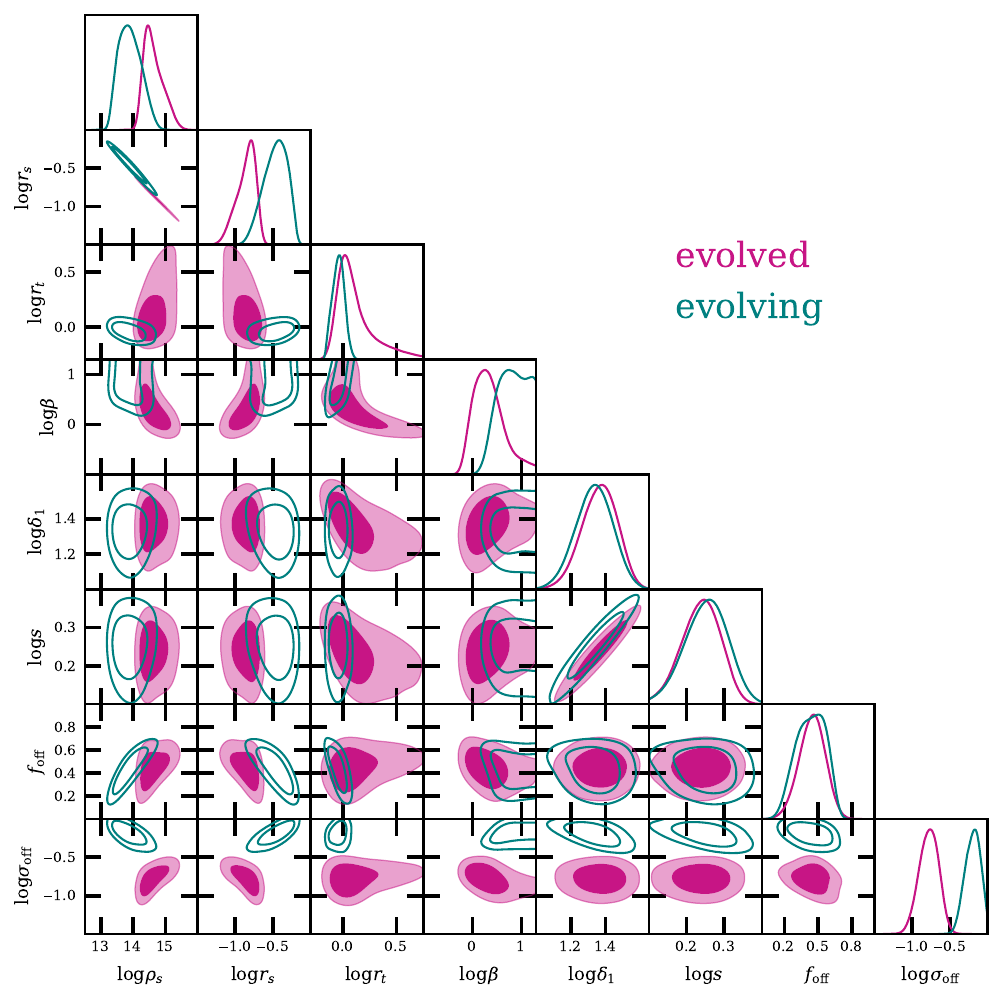}
    \caption{The Preconditioned Monte Carlo algorithm \textsc{pocoMC} \protect\citep{Karamanis2022a,Karamanis2022b} is used to generate Monte Carlo samples of the density profile model parameters \protect\cite{Diemer2022}, fitted to the lensing profiles of the evolved and evolving samples. Shown here are the $68.3\%$ and $95.5\%$ credible regions.}
    \label{fig:PMC}
\end{figure}

The final constraints on these parameters can be seen in Figure~\ref{fig:PMC}. The largest difference between the two samples is in the values of the mis-centering parameters, $f_{\rm off}$ and $\sigma_{\rm off}$. This difference could reflect the difficulty of correctly identifying central galaxies in the evolving, but may also indicate genuine differences in the mass distribution due to ongoing accretion. In the TNG simulations, we have tested centering lensing profiles on the centre-of-mass as opposed to the most bound particle (normally located within the BCG). We find that evolved and evolving systems do have different central profiles for either choice of centre, suggesting a genuine redistribution of mass around the densest point, rather than just an offset of the centre. 

\begin{figure}
    \centering
    \includegraphics[width=0.5\columnwidth]{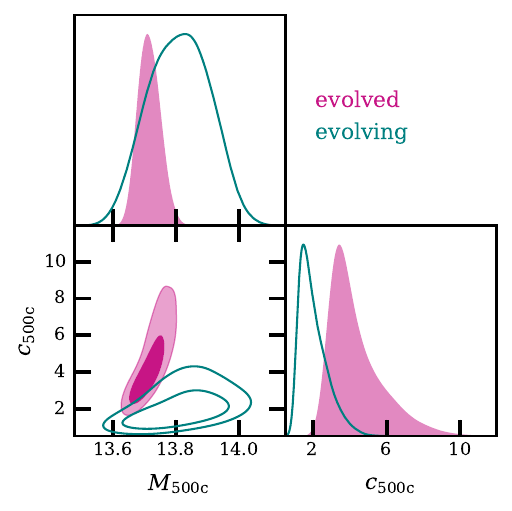}
    \caption{Derived mass and concentration from fits to the weak lensing profiles. The solid and dash-dot lines correspond to the mean values from the catalogue reported masses for the evolved and evolving samples, respectively.}
    \label{fig:M_c}
\end{figure}

\begin{figure}
    \centering
    \includegraphics[width=0.5\columnwidth]{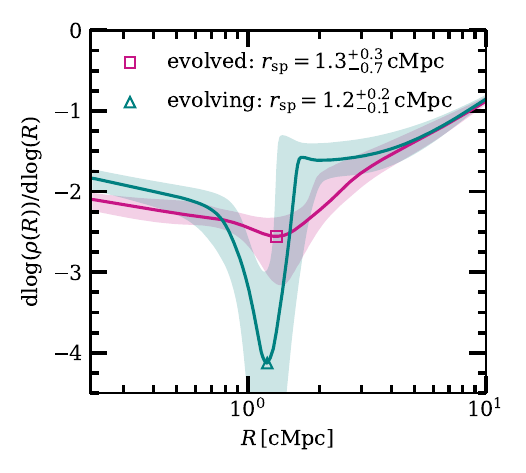}
    \caption{Splashback radii of the evolved and evolving samples. We expect the evolved sample to have a larger splashback radius, for which there is tenuous evidence in our samples.}
    \label{fig:rsp}
\end{figure}

We also see a significant difference in the scale density and radius, $\rho_{\rm s}$ and $r_{\rm s}$. This can be understood as a difference in concentration between these two samples, highlighted in Figure~\ref{fig:M_c}. The evolved sample is much more concentrated, as expected for a sample of clusters that formed earlier and built up larger central densities \citep[e.g.][]{Wechsler2002}. It should be noted that this concentration is not equivalent to an NFW or Einasto concentration --- different density profile definitions lead to different concentrations of up to $\sim\!20\%$ \citep{Diemer2025}. Finally, there is also a marginal difference in the parameters governing the truncation of the Einasto profile, $r_{\rm t}$ (the truncation radius) and $\beta$ (the sharpness of the truncation). The evolving sample has a smaller truncation radius, with a sharper truncation. This suggests that recent infalling matter experienced a shallower potential well, and there has not been sufficient time for an extended splashback region, caused by a growing potential well, to develop. This is evidenced by the splashback radius of the two samples, seen in Figure~\ref{fig:rsp}, which follow the trend found in simulations \citep{More2015,haggar2024sp}, although with large uncertainties in the fitted values.

\begin{figure}
    \centering
    \includegraphics[width=0.9\columnwidth]{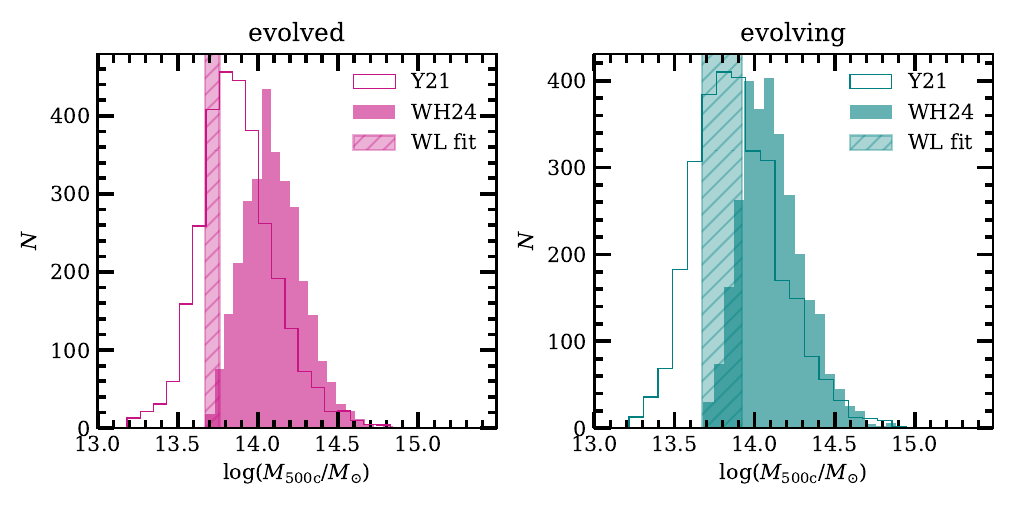}
    \caption{For each sample, we compare the cluster masses taken from either the \citetalias{yang2021} or \citetalias{wen2024} catalogues with the mean mass obtained from our fits to the weak lensing (WL) profiles. Generally, the masses from \citetalias{wen2024} are larger, and those from \citetalias{yang2021} show closer agreement to the lensing masses.}
    \label{fig:mass_comparison}
\end{figure}

Finally, Figure~\ref{fig:mass_comparison} shows a comparison between the weak-lensing obtained masses, and those reported in the two cluster catalogues used in this work, \citetalias{yang2021} and \citetalias{wen2024}. It is clear that the weak-lensing masses show greater agreement with the catalogue masses of \citetalias{yang2021}.

\section{Number density profiles}
\label{app:num_den}

The radial number density profiles of galaxy clusters is often used as an analogous test to estimate the mass distribution in clusters. Here we use the large-scale galaxy information from Sec.~\ref{sec:lss_galaxy_search} to measure the radial number density profiles of the evolved and evolving samples.

\begin{figure*}
    \centering
    \includegraphics[width=0.9\columnwidth]{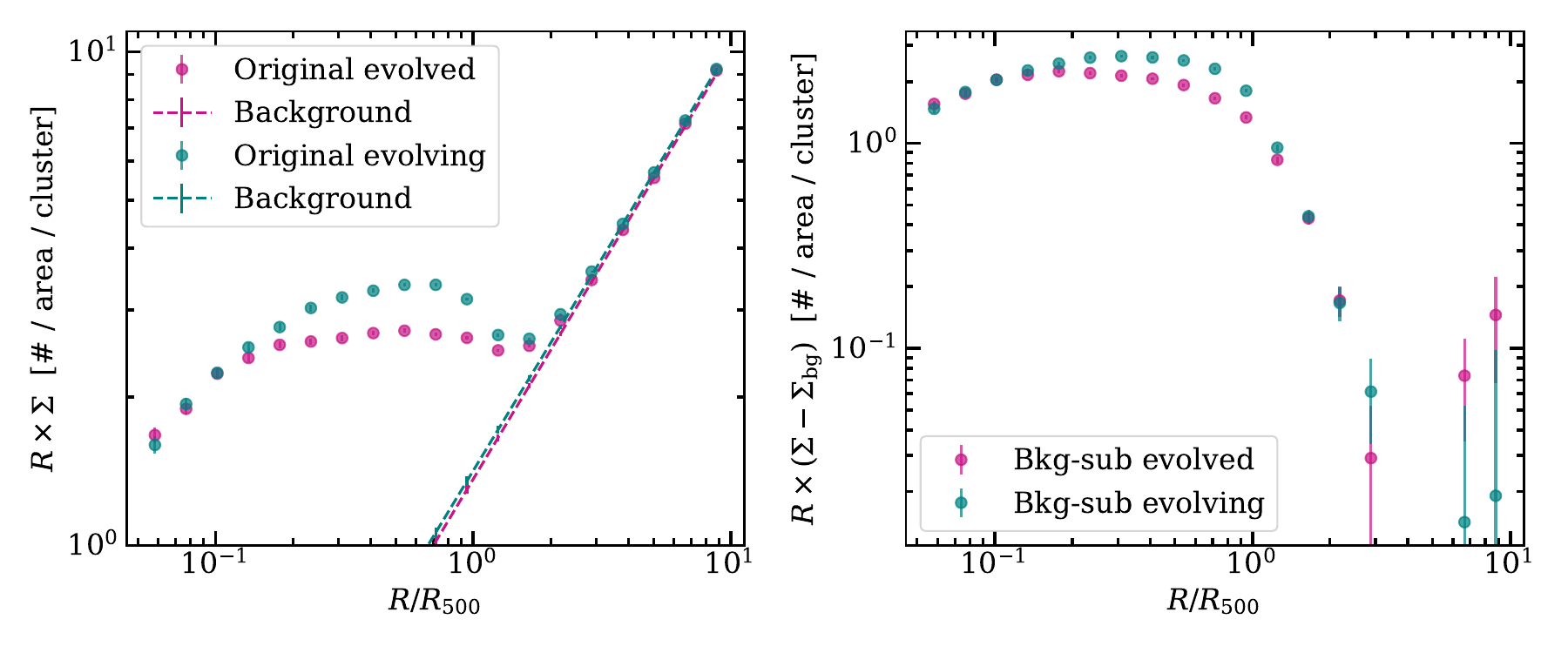}
    \caption{Radial number density profiles of the evolved and evolving samples. The large-scale contribution from non-members is estimated via fitting a power law to the profile at 3-5~Mpc (indicated on the left panel) and subtracting it from the profiles. The subtracted number density profile is on the right panel.}
    \label{fig:num_den_r_ur}%
 \end{figure*}

Figure~\ref{fig:num_den_r_ur} shows the radial number density profiles, measured as the number of galaxies per unit area per cluster in consecutive circular discs centred at the cluster BCGs. The left panel shows the direct measurement, which shows that the number count keeps growing beyond $2R_{500c}$. This is because the galaxies in the outskirts are not necessarily part of the same halo as the cluster. To remove the background contribution, we fit a power law to the outer profiles at $3 - 6 \times R_{500c}$ from the cluster centres (shown by the dashed lines in the left panel) and subtract it from the original measurement. The resulting number density profiles are shown on the right panel. The error bars are measured from 100 bootstrap re-sampling and show the $1-\sigma$ distribution of the measurements. The profiles show that except in the central region, number densities are higher in the evolving sample ($0.3 - 1 \times R_{500c}$), as expected if these systems are experiencing active merging and infall. However, this could also be due to a higher fraction of projected galaxies in these clusters. The relative impact of projection effects will be considered in a future paper (Reid et al.~in prep.).

\bibliography{cl_dynamics_unions}{}
\bibliographystyle{aasjournal}

\end{document}